\documentclass[lettersize,journal]{IEEEtran}
\usepackage[utf8]{inputenc}
\usepackage{cite}
\usepackage[english]{babel}
\usepackage{epsfig}
\usepackage{algorithm}
\usepackage{algorithmic}
\usepackage{amsmath}
\usepackage{amsthm}
\usepackage{amssymb}
\usepackage{dsfont}
\usepackage{subfigure}
\usepackage{balance}
\usepackage{pdflscape}
\usepackage{placeins}
\usepackage{graphicx}
\usepackage{cases}
\usepackage{amsfonts}
\usepackage{multirow}
\usepackage{multicol}
\usepackage{mathrsfs}
\usepackage{color}
\usepackage{balance}
\usepackage{url}
\usepackage{epstopdf}
\usepackage{longtable}
\usepackage{array}
\usepackage{xcolor}
\usepackage{algorithm}
\usepackage{float}
\def\BibTeX{{\rm B\kern-.05em{\sc i\kern-.025em b}\kern-.08em
		T\kern-.1667em\lower.7ex\hbox{E}\kern-.125emX}}
\floatname{algorithm}{Algorithm}


\begin{document}
	\title{A Correlation-Based Design of RIS for Reduced Power Consumption and Simplified Control Circuitry}
	\author{Zina Mohamed, 
		Ammar B. Kouki, and Sonia A\"{i}ssa 
		\thanks{Z. Mohamed is with the Institut National de la Recherche Scientifique (INRS), Montreal, QC, Canada; email: zina.mohamed@inrs.ca. A. B. Kouki is with the \'Ecole de Technologie Sup\'erieure (ETS), Montreal, QC, Canada; email: ammar.kouki@etsmtl.ca. S. A\"{i}ssa is with the Institut National de la Recherche Scientifique (INRS), Montreal, QC, Canada; email: sonia.aissa@inrs.ca}
		\thanks{This work was supported by a Discovery Grant from the Natural Sciences and Engineering Research Council (NSERC) of Canada.}
	}
	\maketitle
	\begin{abstract}
		Aiming at simplifying the hardware structure and reducing the energy consumption in wireless communication via reconfigurable intelligent surfaces (RIS), this paper introduces a novel RIS design founded on the correlation between the phase shift values of the surface elements. First, a correlation analysis is conducted, considering the azimuth angle of a target device within a coverage region spanning from $-80^{\circ}$ to $80^{\circ}$. The correlation is demonstrated for different deployment cases, creating the basis for the new RIS structure, termed Connected-RIS, where correlated elements are designed to share the same control signal. The fundamental performance of the proposed design is then analyzed in terms of control signals, power consumption, and communication system performance, comparing it to two RIS structures with full control: one with the same size as the proposed design, and the other employing the minimum number of elements necessary to satisfy the fair coverage criterion. The correlation-based RIS design enables three-dimensional passive beamforming and significantly reduces the number of required load impedances and control signals, thereby lowering the hardware cost and simplifying the control circuitry. It also achieves substantial power savings as compared to the baseline schemes, while maintaining sufficient gain for a fair radio coverage. For instance, numerical simulations demonstrate that the proposed design reduces the power consumption by almost 86-92\% and the control signals by 83-98\% compared to operation with fully controlled RIS.
	\end{abstract}
	\section{Introduction}
	The RIS (reconfigurable intelligent surface) technology is a key enabler for future wireless networks, thanks to its capability to dynamically control the electromagnetic wave propagation \cite{Chen2024, zhang2024design, Alsabah2021, Zhang2020}. Among multiple ensuing advantages are the improvements in the spectral efficiency, the radio coverage extensions, and the reduced power consumptions compared to conventional relaying techniques \cite{Pei2021}. Recently, with the growing demands for low-power massive connections and green radio in future wireless communications, the reduction of the energy consumption in RIS-assisted communication systems has been gaining increasing research interest \cite{Mo2020,Mohamed2020}.
	
	From the architecture viewpoint, conventional RIS structures are typically composed of a number of nearly passive elements, such as tunable meta-atoms, PIN diodes, varactors, and liquid crystals \cite{Gros2021}. Although each unit is energy-efficient on its own, the overall hardware complexity and power consumption increase significantly as the surface size increases or that multiple RISs are deployed within the wireless network. Several studies and measurements have shown important findings in this regard. For instance, \cite{Zubair2024} showed that while a PIN diode only consumes 0.06 mW, the supporting components increase this to 3.7 mW per element. As the number of RIS elements increases, so does the power consumption. In \cite{Zubair2024}, it was shown that a typical 1-bit RIS of size $27 \times 28$ consumes an average power of 2.81 W, with micro-controllers using about 160 mW each, and the logic circuit consuming 1.8 W in total. According to \cite{li2023enhan}, deploying ten RIS units, each with 1024 elements, can waste over 35 W of static power just from the control circuits. Also, a 2-bit RIS with 256 elements consumes about 153 W when the total radiated power is 64 W \cite{Dai2020}, and $33 \times 33$ elements operating at 5.8 GHz consume approximately 1 W \cite{Pei2021}. At high frequencies, where the sizes of unit cells are smaller, the power consumption can be even more significant, e.g., about 200W/${\rm m}^2$ at an operating frequency of 10.6 GHz \cite{Zubair2024}.
	
	While specific prices are not available in the open literature, it is clear that the RIS fabrication cost will be significant, especially as the number of elements increases. Indeed, the more elements a RIS comprises, the higher the number of control signals that are required to configure the surface would be. For instance, an element offering $2^n$ states requires $n$ bits for the control. With $N^2$ reflecting elements in a RIS, an equivalent number of control lines would be needed to manage them independently \cite{Saggese2024}. This scaling can quickly lead to a substantial increase in both the physical wiring and the logical control complexity, as each unit-cell, often implemented with PIN diodes, demands its own dedicated control signal \cite{Rafique2023}. As the RIS grows in size, the micro-controller or FPGA responsible for generating these signals must also scale, which necessitates additional interface circuitry, hence increasing the overall design cost. To mitigate this, unit-cell grouping can be employed, where several elements are to controlled via shared control lines \cite{Liu2022}. While this approach can reduce the number of required control paths, it comes at the expense of a reduced flexibility and ability in generating complex radiation patterns, hence a trade-off between the simplicity of the control circuitry and the performance of the RIS.
	
	The high power consumption and the need for sophisticated control mechanisms contribute to the overall cost, making it a critical factor in the widespread adoption of this promising technology. To reduce the implementation cost and optimize the functioning of these intelligent surfaces, it is crucial to design simplified structures for the RIS and its corresponding control circuit. The goal is to enable the implementation of a large aperture while maintaining efficiency and scalability. Indeed, an optimized RIS design is expected to not only minimize the hardware complexity and fabrication cost, but also lower the power consumption, thereby enhancing the overall sustainability of RIS-assisted wireless systems.
	
	This is exactly where this paper's contribution becomes significant, as it consists in advancing an innovative solution that simplifies the RIS architecture while ensuring seamless control and adaptability, ultimately facilitating large-scale deployments of RISs in next-generation wireless networks. Specifically, the paper proposes a new RIS design founded on the correlation between the phase shift values of the surface elements. In so doing, several contributions are advanced:
	\begin{itemize}
		\item A methodology is proposed for analytically determining the minimum RIS size required to achieve fair coverage, defined by equalizing the power received at the user-equipment (UE) through the RIS-assisted path and the one that would result when the direct communication path between the base station (BS) and the UE exists. The process involves analyzing the received powers, identifying the necessary gain margins, and finding the minimum number of RIS elements essential for effective operation, while considering realistic scenarios based on 3GPP specifications \cite{3GPP2018}.
		\item The required RIS sizes for various deployment cases, based on the RIS position relative to the BS and the UE, are determined using the proposed methodology, considering two gain margin values, namely, 3 dB and 6 dB.
		\item Using the obtained size values, phase-shift matrices are generated for different steering directions within the coverage region spanning from $-80^{\circ}$ to $80^{\circ}$ in the azimuth plane. Then, a correlation analysis is performed to determine the correlated elements, and the level of the correlation.
		\item Based on results from the above, a novel RIS design is proposed, in which the correlated elements are connected together and share the same control signal.
		\item The main properties of the proposed Connected-RIS design, such as the number of load impedances, the number of control signals, the power consumption, the achieved RIS gain, and the resulting data rate when the Connected-RIS is integrated into a communication system, are thoroughly evaluated. The results show that the proposed design significantly reduces the number of required load impedances and direct current (DC) control lines, the number of codewords, and the overall power consumption, leading to a substantial decrease in hardware complexity and implementation cost. For instance, the Connected-RIS reduces the overall power consumption by approximately 86-92\% and the control signals by 83-98\% as compared to the fully controlled RIS designs.
		\item Finally, a case study is conducted to evaluate the number of codewords required to cover a specific area within an azimuth range spanning from $-80^{\circ}$ to $80^{\circ}$. Based on the ensuing results, two deployment approaches are proposed to achieve the desired coverage: a dynamically configured RIS, and a fixed multi-RIS configuration. These two configurations are then assessed in terms of power consumption and the achievable gain when deploying the Connected-RIS architecture.
	\end{itemize}
	
	{\it Notations}: Vectors and matrices are denoted by lower- and upper-case boldface letters, respectively. The matrix operators $(\cdot)^H$ and $(\cdot)^T$ refer to the Hermitian transpose and transpose, respectively. The symbol $\lVert. \lVert$ denotes the Euclidean norm, and $\lvert. \lvert$ is the modulus of a complex number. The mathematical expectation is denoted by $\mathbb{E}[\cdot]$.
	\section{The RIS-aided Communication Model}
	\label{Section-SystemModel}
\begin{figure}[h!]
		\centering
		\includegraphics[width=1.07\linewidth]{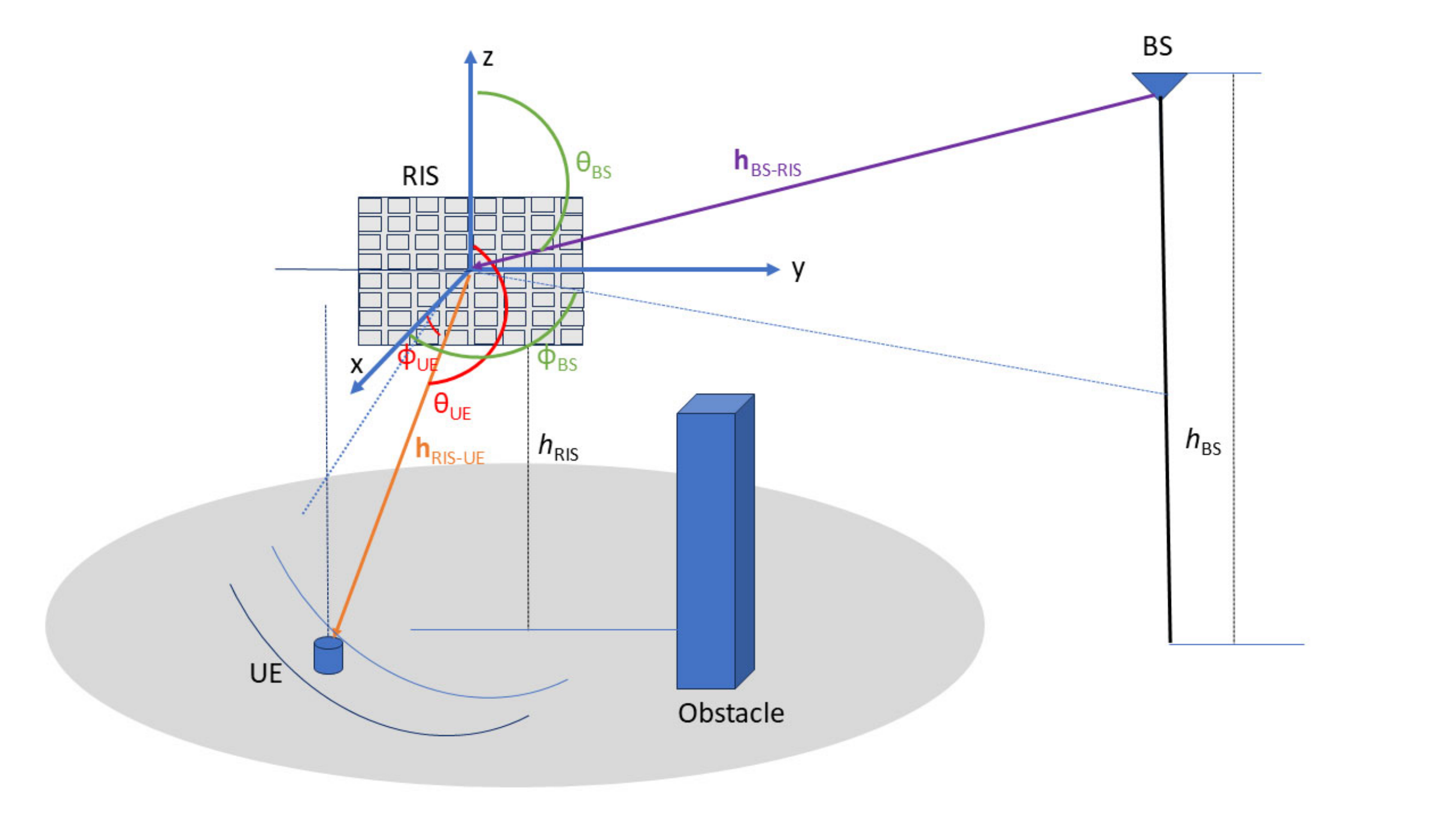}
		\caption{Illustration of the RIS-aided downlink communication system.}
		\label{SystemModel}
	\end{figure}

	We consider the RIS-aided wireless communication system illustrated in Fig.~\ref{SystemModel}. The RIS, deployed in the $x-z$ plane of the coordinate system shown, is centered at the origin (0,0,0). The RIS consists of a uniform planar array (UPA) of $N_y$ columns and $N_z$ rows. For simplicity, we assume that $N_{\rm y}=N_{\rm z}$. The width and the length of each element are denoted by $d_{\rm y}$ and $d_{\rm z}$, respectively. The size of the $n^{\rm th}$ element is hence given by $\mathcal{A}_{n}=d_{\rm y}d_{\rm z}$. The spacing between elements is denoted by $\delta_1$. Let us denote by $\mathbf{\Psi}=\left(\mathbf{\psi}_1,\cdots,\mathbf{\psi}_{N_{\rm y}}\right)\in \mathbb{C}^{N_{\rm y} \times N_{\rm z}}$ the matrix of the RIS reflection coefficients, where $\mathbf{\psi}_n=\left[\beta_{n,1} e^{j \phi_{n,1}},\cdots,\beta_{n,N_{\rm z}} e^{j \phi_{n,{N}_{\rm z}}}\right] \in \mathbb{C}^{N_{\rm z}\times 1},~\phi_{n,m} \in [0,2\pi)$ and $\beta_{n,m} \in [0,1],$ $\forall m\in\{1,\cdots,N_{\rm y}\}$ and $~\forall n\in\{1,\cdots,N_{\rm z}\}$. Also, let $N$ denote the total number of elements in the RIS.
	
	We assume that the BS, located at a distance $d_{\rm BS,RIS}$ from the center of the RIS, is characterized by an azimuth angle $\phi_{\rm BS}$ and an elevation angle $\theta_{\rm BS}$. The single-antenna UE is located at a distance $d_{\rm RIS,UE}$ from the center of the RIS, and characterized by its azimuth and elevation angles denoted by $\phi_{{\rm UE}}$ and $\theta_{{\rm UE}}$, respectively. The direct line-of-sight (LoS) link between the BS and the UE is assumed to be unavailable due to blockage.
	
	Let $\mathbf{h}_{\rm BS,RIS}=[h_{{\rm BS},1},\cdots,h_{{\rm BS},N}] \in \mathbb{C}^{1\times N}$ and $\mathbf{h}_{\rm RIS,UE}=[h_{1,{\rm UE}},\cdots,h_{N,{\rm UE}}]\in \mathbb{C}^{N\times 1}$ represent the channel vectors between the BS and the RIS, and between the RIS and UE, respectively. Without loss of generality, the channels are modelled by considering the effects of the large-scale fading, represented by $L(d)$, and the small-scale fading, indicated by $\tilde{h}$.\footnote{This modeling is essential for understanding the performance of the RIS in real-world applications, where factors such as distance, path loss, and fading can significantly influence the signal integrity.}
	
	For the large-scale fading, the 3GPP UMi-Street Canyon path loss model \cite[Page 26]{3GPP20181} is used.\footnote{This model provides precise predictions for signal coverage and path loss, making it crucial for network planning and performance evaluation in modern wireless communication systems, including 5G and B5G networks, as it accurately reflects the unique propagation characteristics of dense city centers with tall buildings.} Hence, we can write
	\begin{align}
		L^{\rm dB}(d) = 32.4+21 \log_{10}(d) + 20 \log_{10}(f),
	\end{align}
	\noindent where $d$ is the distance in meters (m), and $f$ is the center frequency normalized by 1 GHz. All distance related values are normalized by 1 m.
	
	The small-scale fading coefficients follow a Rician distribution \cite{3GPP20181}. Therefore, the channel coefficients between the BS antenna and the RIS can be expressed as
	\begin{align}
		\mathbf{h}_{{\rm BS,RIS}}&=\sqrt{L(d_{\rm BS,RIS})} \tilde{\mathbf{h}}_{{\rm BS,RIS}}\nonumber\\
		&=\sqrt{L(d_{\rm BS,RIS})}\Bigg(\sqrt{\frac{\kappa_{{\rm BS,RIS}}}{\kappa_{{\rm BS,RIS}}+1}}\mathbf{h}_{{\rm BS,RIS}}^{\rm L}\nonumber\\&\qquad\qquad\qquad~~+\sqrt{\frac{1}{\kappa_{{\rm BS,RIS}}+1}}\mathbf{h}_{{\rm BS,RIS}}^{\rm NL}\Bigg),
		\label{Chan1}
	\end{align}
	\noindent where $\kappa_{{\rm BS,RIS}}$ is the Rician factor, $L(d_{\rm BS,RIS})$ is the path loss component in linear scale, with $d_{\rm BS,RIS}$ being the Euclidean distance between the BS and the RIS, and where $\mathbf{h}_{{\rm BS,RIS}}^{\rm L}$ and $\mathbf{h}_{{\rm BS,RIS}}^{\rm NL}$ respectively denote the deterministic LoS component and the random non-line-of-sight (NLoS) component of the channel from the BS to the RIS. The LoS component is expressed as
	\begin{equation}	
		\mathbf{h}_{{\rm BS,RIS}}^{\rm L}=\left[1,\cdots,\exp\left(\frac{j2\pi \delta_1 (N-1) \mathcal{\xi}_1}{\lambda}\right)\right],
	\end{equation}
	\noindent where $\lambda$ is the carrier wavelength, $\mathcal{\xi}_1=\cos(\theta_{\rm UE}-\theta_{\rm BS})\sin(\phi_{\rm UE}-\phi_{\rm BS})+\sin(\theta_{\rm UE}-\theta_{\rm BS})$, and $\delta_1$ is the spacing between the RIS elements, set to $\lambda/8$ \cite{Estakhri2016}, \cite{app122211857}. For the NLoS component $\mathbf{h}_{{\rm BS,RIS}}^{\rm NL}=[h_{{\rm BS},1}^{\rm NL},\cdots,h_{{\rm BS},N}^{\rm NL}]$, the entries are modeled as i.i.d. circularly symmetric complex Gaussian random variables with zero mean and unit variance.
	
	Similarly, the channel between the RIS and the UE is expressed as
	\begin{align}
		\mathbf{h}_{{\rm RIS,UE}}&=\sqrt{L(d_{\rm RIS,UE})}\tilde{\mathbf{h}}_{{\rm RIS,UE}}\nonumber\\
		&=\sqrt{L(d_{\rm RIS,UE})}\left(\sqrt{\frac{\kappa_{{\rm RIS,UE}}}{\kappa_{{\rm RIS,UE}}+1}}\mathbf{h}_{{\rm RIS,UE}}^{\rm L}\right.\nonumber\\&\qquad\qquad\qquad\left.+\sqrt{\frac{1}{\kappa_{{\rm RIS,UE}}+1}}\mathbf{h}_{{\rm RIS,UE}}^{\rm NL}\right),
		\label{Chan2}
	\end{align}
	\noindent where $\kappa_{{\rm RIS,UE}}$ is the link's Rician factor, $L(d_{\rm RIS,UE})$ is the path-loss component, and $d_{\rm RIS,UE}$ is the Euclidean distance between the RIS and the UE. The LoS component of the channel, i.e., $\mathbf{h}_{{\rm RIS,UE}}^{\rm L}$, is expressed as
	\begin{align}
		\mathbf{h}_{{\rm RIS,UE}}^{\rm L}=\left[1, \cdots,\exp\left(\frac{j2\pi \delta_1 (N-1) \mathcal{\xi}_2}{\lambda}\right)\right],
	\end{align}
	\noindent where $\mathcal{\xi}_2=\cos(\theta_{\rm UE})\sin(\phi_{\rm UE})+\sin(\theta_{\rm UE})$.
	\section{Coverage Dependencies on the RIS Size}
	\subsection{RIS Size for Fair Coverage}
	This section presents a methodology for determining the minimum number of RIS elements needed to ensure fair UE coverage in the communication system described above, specifically examining the RIS size that would be needed to provide the UE with the required service level via the RIS-assisted path, e.g., for users in dead zones or at the cell edge, as compared to the case when the UE can be served directly by the BS in case a LoS in between exists. By calculating the required RIS size to make the power received via the RIS-assisted path match or exceed that of a direct BS-UE link, the approach addresses signal disparities and ensures balanced performance for all users regardless of location.
	
	Firstly, let us denote by $h_{\rm BS, UE} \in \mathbb{C}^{1\times 1}$ the channel coefficient between the BS and the UE, given by
	\begin{align}
		h_{{\rm BS,UE}}&=\sqrt{L(d_{\rm BS,UE})}\tilde{h}_{{\rm BS,UE}}\nonumber\\
		&=\sqrt{L(d_{\rm BS,UE})}\Bigg(\sqrt{\frac{\kappa_{{\rm BS,UE}}}{\kappa_{{\rm BS,UE}}+1}}h_{{\rm BS,UE}}^{\rm L}\nonumber\\&\qquad\qquad\qquad+\sqrt{\frac{1}{\kappa_{{\rm BS,UE}}+1}}h_{{\rm BS,UE}}^{\rm NL}\Bigg),
		\label{Chan1}
	\end{align}
	\noindent where $\kappa_{{\rm BS,UE}}$ is the Rician factor of the direct link (BS-UE), $L(d_{\rm BS,UE})$ is the path-loss component, in which $d_{\rm BS,UE}$ denotes the Euclidean distance between the BS and the UE, and where $h_{{\rm BS,UE}}^{\rm L}$ and $h_{\rm BS,UE}^{\rm NL}$ are the deterministic LoS component and the random NLoS component, respectively.	
	
	Next, we outline the methodology for determining the minimum RIS size (number of elements) that is required for the RIS to ensure that the received power via the RIS-assisted path is at least equal to that of the direct path between the BS and the UE in the absence of blockage. This ensures fair coverage for users in dead zones or at cell edges, providing them with signal quality comparable to users with direct LoS connections to the BS.
	\subsubsection{Received Power}	
	The UE's received power via the direct path (BS-UE) can be expressed as
	\begin{align}
		P_{\rm UE}^{\rm D}
		=P_{\rm t} L\left(d_{\rm BS,UE}\right)\left\lvert\tilde{h}_{\rm BS,UE}\right\rvert^2.
		\label{eq1}
	\end{align}
	The UE's received power from the indirect path (BS-RIS-UE) is given by
	\begin{align}
		&P_{\rm UE}^{\rm I}=P_{\rm t} G_{\rm RIS} L\left(d_{\rm BS,RIS}\right)L\left( d_{\rm RIS,UE}\right)\left\lvert\tilde{\mathbf{h}}_{\rm BS,RIS}\right\rvert^2\left\lvert\tilde{\mathbf{h}}_{{\rm RIS,UE}}\right\rvert^2,
		\label{eq2}
	\end{align}
	\noindent where $G_{\rm RIS}$ is the gain of the RIS.
	\subsubsection{Fair Coverage Requirement}	
	To determine the minimum RIS size that guarantees equal received power between the RIS-assisted path and the direct path from the BS, we need to consider several factors and establish the fair coverage requirement condition, which can be satisfied by the following equation:
	\begin{align}
		P_{\rm UE}^{\rm I}\geq P_{\rm UE}^{\rm D}.
		\label{eq}
	\end{align}
	By replacing Eq.~(\ref{eq1}) and Eq.~(\ref{eq2}) into Eq.~(\ref{eq}), and then rearranging the inequality, the RIS gain satisfying the coverage requirement can be expressed as
	\begin{align}
		G_{\rm RIS}&\geq\frac{L\left(d_{\rm BS,UE}\right)\left\lvert\tilde{h}_{\rm BS,UE}\right\rvert^2}{L\left(d_{\rm BS,RIS}\right)L\left( d_{\rm RIS,UE}\right)\left\lvert\tilde{\mathbf{h}}_{\rm BS,RIS}\right\rvert^2\left\lvert\tilde{\mathbf{h}}_{{\rm RIS,UE}}\right\rvert^2}.
	\end{align}
	To determine the minimum RIS gain in linear scale, denoted by $G_{\rm RIS, min}$, the received power through the cascaded link should be at least equal to the received power through the direct link.
	
	Considering the path loss and small-scale fading, along with the previous inequality, the minimum RIS gain satisfying the fair coverage condition is given, in linear scale, by
	\begin{align}
		G_{\rm RIS, min}=\frac{L\left(d_{\rm BS,UE}\right)\left\lvert\tilde{h}_{\rm BS,UE}\right\rvert^2}{L\left(d_{\rm BS,RIS}\right)L\left(d_{\rm RIS,UE}\right)\left\lvert\tilde{\mathbf{h}}_{\rm BS,RIS}\right\rvert^2\left\lvert\tilde{\mathbf{h}}_{{\rm RIS,UE}}\right\rvert^2}.
		\label{gmin}
	\end{align}
	\subsubsection{Incorporating a Gain Margin}
	After obtaining the minimum RIS gain satisfying the fair coverage condition and transforming it to the dB scale, $G_{\rm RIS, min}^{\rm dB}$, a gain margin $\Delta G^{\rm dB}$ is added to the latter to obtain the required gain, denoted as $G_{\rm req}$. This gain margin serving as a proactive compensation accounts for the effect of the phrase-shift quantization based one the correlation.\footnote{Hereafter, $\Delta G^{\rm dB}$ is set to 3 or 6 dB.} Hence, $G_{\rm req}$ can be expressed as
	\begin{align}
		G_{\rm req}^{\rm dB}=G_{\rm RIS, min}^{\rm dB}+\Delta G^{\rm dB}.
	\end{align}
	\subsubsection{Determining the Required RIS Size}	
	As the RIS gain is proportional to the total number of elements squared, the gain $G_{\text{req}}$, in linear scale, is proportional to $N^2$. Therefore, the minimum required number of RIS elements to satisfy this gain can be obtained as follows:
	\begin{align}
		N = \sqrt{G_{\text{req}}}.
	\end{align}
	
	Given the planar structure of the RIS, as described in section II, we assume that its elements are arranged in a square configuration. Consequently, the total number of elements, $N$, is distributed as $N_{\rm y}$ and $N_{\rm z}$, with $N_{\rm y}$ denoting the number of rows and $N_{\rm z}$ denoting the number of columns. The angular coverage will depend on the specific phase-shift matrix used. Operating at frequency $f$, the half-power beamwidth is approximated as $\theta = \arcsin\left(\frac{\lambda}{D}\right)$, where $D=N_{\rm z} \delta_1$. Recall that $\delta_1$ is the element spacing.
	\subsection{Impact of the RIS Position}
\begin{figure}
		\centering
		\includegraphics[width=3.5in,height=2.5in]{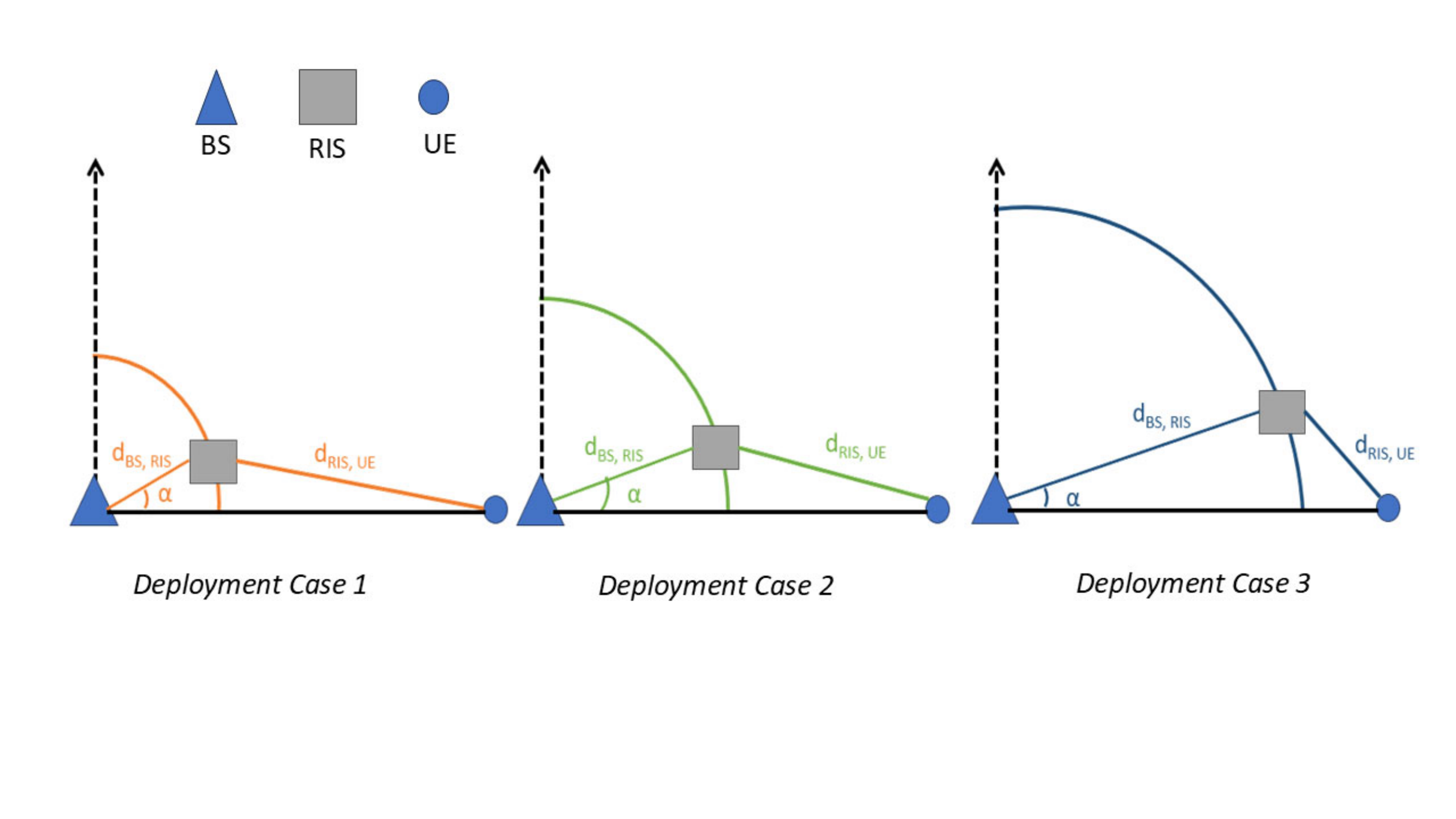}
		\vspace*{-0.7in}
		\caption{Illustration of the deployment cases.}
		\label{deploy1}
	\end{figure}

	Next, building upon the methodology established earlier, we analyze the impact of the RIS position on the minimum RIS size for fair coverage. By applying the derived formula across various deployment cases (cf. Fig.~\ref{deploy1}), where the RIS is positioned at different locations relative to the BS and the UE, this analysis reveals the spatial sensitivity of the RIS design and offers practical insights into optimal deployment strategies.
	
	Three deployment cases are considered in this study: 1) {\it Deployment Case 1}: the RIS is positioned near the BS, while satisfying the far-field condition; 2) {\it Deployment Case 2}: the RIS is located midway between the BS and the UE; and 3) {\it Deployment Case 3}: the RIS is assumed to be in close proximity to the UE. Using the formula relating the distances and the angle $\alpha$ formed by the BS-RIS link and the direct link, given by $d_{\rm RIS, UE}=\sqrt{d_{\rm BS, UE}^2+d_{\rm BS, RIS}^2-2d_{\rm BS, UE}d_{\rm BS,RIS}\cos(\alpha)}$, the following figures are generated.
	
	\begin{figure}[h!]
		\centering
		\hspace*{-0.2in}
		\includegraphics[width=4in,height=3.2in]{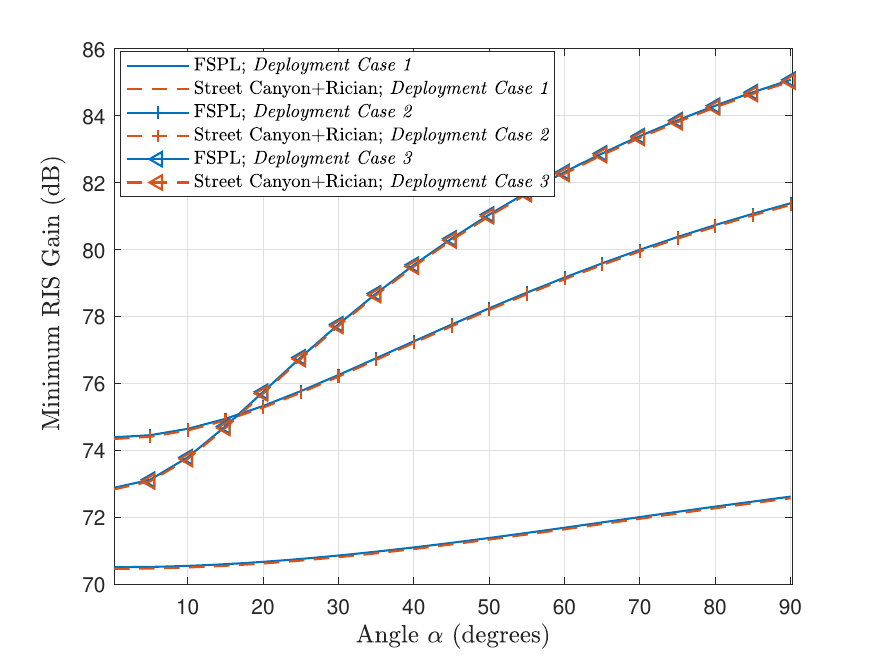}
		\caption{The minimum RIS gain versus the angle $\alpha$.}
		\label{DC1222-a}
	\end{figure}
	
	Figure \ref{DC1222-a} illustrates the minimum RIS gain (in dB) as a function of the angle $\alpha$ (in degrees) for the deployment cases described above. The free-space path loss (FSPL), calculated in dB based on $L^{\rm dB}(d)= 92.45+20 \log_{10}(d)+ 20 \log_{10}(f)$, with $d$ being the distance between the transmitter and the receiver in Km and $f$ denoting the carrier frequency in GHz, is used for comparison. Additionally, the Street Canyon and Rician fading models, as described in section II, and Eqs. (1)–(4), are also considered for benchmarking. As expected, the minimum RIS gain increases as the distance between the BS and the RIS increases, reaching its maximum at $\alpha=90^\circ$, which is a straightforward result of the increased path loss. We also observe that the system operating under FSPL conditions consistently requires slightly higher gain compared to operation under Street Canyon conditions at all evaluated distances. The periodic variation of the gain with the angle indicates that the RIS effectiveness depends on its distance w.r.t. the BS and the UE. These results highlight how the RIS performance is affected by both the propagation environment and its position w.r.t. the source and the destination.
	
	\begin{figure}[h!]
		\centering
		\hspace*{-0.2in}
		\includegraphics[width=4in,height=3.2in]{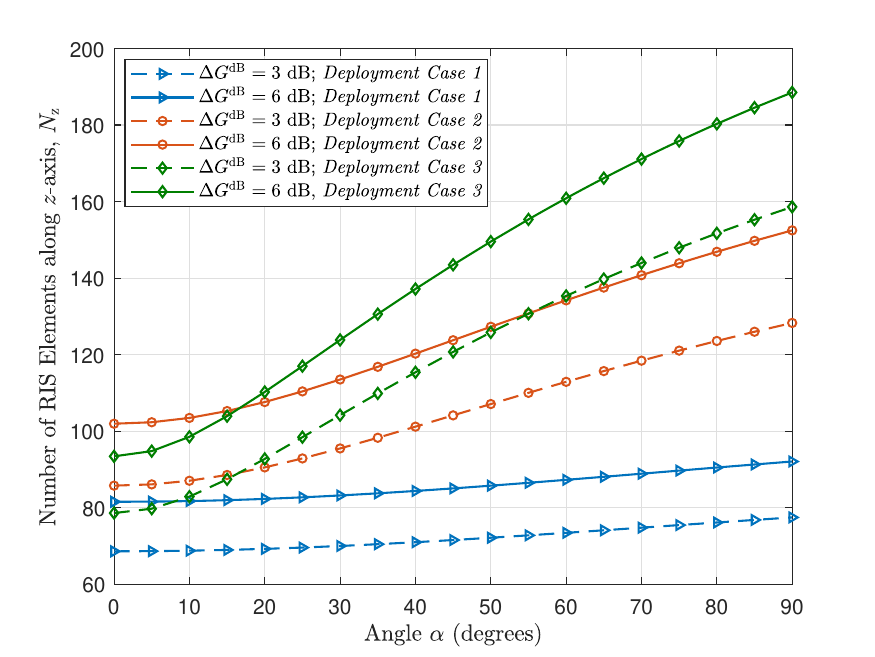}
		\caption{The required number of elements along $z$-axis, $N_z$. Here, $d_{\rm BS, UE}=100$ m.}
		\label{DC1222-b}
	\end{figure}
	
	In Fig.~\ref{DC1222-b}, the required number of RIS elements, $N_{\rm z}$, is plotted for the same deployment cases.\footnote{For clarity of the figure, the curves are plotted as a function of $N_{\rm z}$, the number of RIS columns. Since the RIS is square-shaped, the behavior with respect to $N_{\rm z}$ is representative of that with respect to $N_{\rm y}$, i.e., the number of rows, and thus the total number of elements.} The plots show how $N_{\rm z}$ varies with the angle $\alpha$, factoring in different RIS gain margins and distances between the BS, the RIS, and the UE. As the angle increases, the required number of elements also rises, reflecting an increase in the distance between the RIS and the UE. Lower RIS gain margins, e.g., 3 dB, necessitate less elements compared to higher gain margins, e.g., 6 dB, especially at larger angles. Additionally, as the distance between the RIS and the UE increases, e.g., from 20 m, to 50 m, to 70 m, the number of required elements increases significantly, with sharper growth at higher angles. This demonstrates that the RIS gain margin and the RIS position both play a crucial role in determining the number of required elements to achieve fair coverage.
	
	\begin{table*}[h]
		\centering
		\caption{System Parameters and RIS Design Characteristics for Different Deployment Cases. Here, The Surface Structure is the Conventional Fully Controlled RIS}
		\label{tab:deployment}
		\begin{tabular}{|l||c|c|c|}
			\hline
			\textbf{Parameters} & \textbf{\textit{Deployment Case 1}} & \textbf{\textit{Deployment Case 2}} & \textbf{\textit{Deployment Case 3}} \\
			\hline\hline
			$d_{\rm BS,UE}$ (m) & 100 & 100 & 100 \\
			$d_{\rm BS,RIS}$ (m) & 20 & 50 & 70 \\
			$d_{\rm RIS,UE}$ (m) & 81.5 & 55.7 & 47.0 \\
			\hline
			$G_{\rm RIS,min}$ (dB) & 70.6250 & 75.3370 & 75.6557 \\
			Min-Gain RIS Size & $57\times57$ & $76\times76$ & $78\times78$ \\
			\hline
			$G_{\rm req}^{3\rm dB}$ (dB) & 73.6250 & 78.3370 & 78.6557 \\
			3dB-RIS Size & $70\times70$ & $91\times91$ & $93\times93$ \\
			\hline
			$G_{\rm req}^{6\rm dB}$ (dB) & 76.6250 & 81.3370 & 81.6557 \\
			6dB-RIS Size & $83\times83$ & $108\times108$ & $110\times110$ \\
			\hline
		\end{tabular}
	\end{table*}
	
	Considering the system model described in section II, operating in the mid-band spectrum at 5 GHz, using realistic values for the frequency and the BS's transmit power as specified in \cite{3GPP2018}, next, we apply the methodology to calculate the minimum number of RIS elements that is required in the deployment cases described above. The system parameters for the three deployment cases are summarized in Table \ref{tab:deployment}. In addition, the Rician factors incorporated in Eqs. (2), (4) and (6) are set as follows: $\kappa_{{\rm BS,RIS}}^{\rm dB} = 10$ dB, $\kappa_{{\rm RIS,UE}}^{\rm dB} = 10$ dB, and $\kappa_{{\rm BS,UE}}^{\rm dB} = 1$ dB.
	\subsubsection{{\it Deployment Case 1}}
	Using the Street Canyon model and the system parameters defined in Table \ref{tab:deployment}, the path losses can be obtained as follows: $L^{\rm dB}d_{\rm BS, UE}=87.07$ dB, $L^{\rm dB}d_{\rm BS, RIS}=72.4$ dB and $L^{\rm dB}d_{\rm RIS, UE}=85.30$ dB. The channel power gain, in linear scale, can be computed using the expression $\lvert h\rvert^2 =10^{-L^{\rm dB}d/10}\left(\frac{\kappa}{\kappa+1} + \frac{1}{\kappa+1}\right)$, where $\kappa \in\{\kappa_{{\rm BS,RIS}}, \kappa_{{\rm RIS,UE}}, \kappa_{{\rm BS,UE}}\}$ denotes the Rician factor in linear scale. Using this formula, we obtain the following results: $\lvert \mathbf{h}_{\rm BS, RIS}\rvert^2=$ $10^{-L^{\rm dB}d_{\rm BS, RIS}/10}\left(\frac{\kappa_{{\rm BS,RIS}}}{\kappa_{{\rm BS,RIS}}+1} + \frac{1}{\kappa_{{\rm BS,RIS}}+1}\right)=$ $5.7544~10^{-8}$, $\lvert \mathbf{h}_{\rm RIS, UE}\lvert^2=10^{-L^{\rm dB}d_{\rm RIS, UE}/10}\left(\frac{\kappa_{{\rm RIS,UE}}}{\kappa_{{\rm RIS,UE}}+1} + \frac{1}{\kappa_{{\rm RIS,UE}}+1}\right)=$ $2.9512~10^{-9}$, and $\lvert h_{\rm BS, UE}\lvert^2=$ $10^{-L^{\rm dB}d_{\rm RIS, UE}/10} \left(\frac{\kappa_{{\rm BS,UE}}}{\kappa_{{\rm BS,UE}}+1} + \frac{1}{\kappa_{{\rm BS,UE}}+1}\right)=$ $1.9596\times 10^{-9}$. Therefore, the minimum RIS gain, based on Eq. (\ref{gmin}), is equal to $1.1539~10^{7}$, which is equivalent to $70.625$ dB. To achieve this gain, the minimum RIS configuration, referred to as min-gain RIS in this work, consists of $57 \times 57$ elements.
	\begin{itemize}
		\item \textbf{Design Characteristics with 3 dB Gain Margin:} To satisfy a gain margin of 3 dB, the required gain becomes $G_{\rm req}^{\rm dB}$=73.625 dB. This corresponds to at least $N^2\geq4799$ elements, achieved with a square structure of $70 \times 70$ elements. With an operating frequency of 5 GHz and an inter-element spacing of $\lambda/8=0.0075$ m, the RIS edge length is $D=0.0075 \times 16= 52.5$ cm, resulting in a surface of size 52.5 cm by 52.5 cm and a half-power beamwidth of approximately $\theta=6.5624^{\circ}$.
		\item \textbf{Design Characteristics with 6 dB Gain Margin:} For a 6 dB margin, the required gain rises to $G_{\rm req}^{\rm dB}$=76.625 dB. This can be achieved with a configuration of $83 \times 83$ elements, giving a surface of size 62.25 cm by 62.25 cm and a beamwidth of $\theta=5.5319^{\circ}$.
	\end{itemize}
	\subsubsection{{\it Deployment Case 2}}
	In this case, the minimum RIS gain is $G_{\rm RIS, min} = 75.3370$ dB, corresponding to a min-gain RIS of $76 \times 76$ elements.	
	\begin{itemize}
		\item \textbf{Design Characteristics with 3 dB Gain Margin:} When accounting for a 3 dB margin, the gain increases to $G_{\rm req}^{\rm dB}=78.3370$ dB, which necessitates a RIS of $91 \times 91$ elements. The size becomes 68.25 cm by 68.25 cm, and the beamwidth is approximately $\theta=5.04^{\circ}$.
		\item \textbf{Design Characteristics with 6 dB Gain Margin:} In this setup, the required gain is $G_{\rm req}^{\rm dB}=81.3370$ dB, met with $108 \times 108$ elements, which results in a RIS size of 81 cm by 81 cm and a narrower beamwidth of $\theta=4.2399^{\circ}$.
	\end{itemize}
	\subsubsection{{\it Deployment Case 3}}
	In this case, the required gain is $G_{\rm RIS, min}= 75.6557$ dB, leading to a min-gain RIS of $78 \times 78$ elements.
	\begin{itemize}
		\item \textbf{Design Characteristics with 3 dB Gain Margin:} By adding a 3 dB margin, the required gain is $G_{\rm req}^{\rm dB}= 78.6557$ dB, satisfied with a square RIS of $93 \times 93$ elements. The resulting size is 69.75 cm by 69.75 cm, and the beamwidth is approximately $4.9274^{\circ}$.
		\item \textbf{Design Characteristics with 6 dB Gain Margin:} For a 6 dB margin, a gain of 81.6557 dB is required, corresponding to a configuration of $110 \times 110$ elements. The RIS in this case spans 82.5 cm by 82.50 cm and has a beamwidth of about $4.1826^{\circ}$.
	\end{itemize}
	\section{The Connected-RIS Design}
	Now, we assume that $\theta_{\rm BS}$, $\phi_{\rm BS}$ and $\theta_{\rm UE}$ are fixed, and change the position of the UE w.r.t. the azimuth angle $\phi_{\rm UE}$.\footnote{In urban environments, the main challenge for the RIS deployments is ensuring coverage in the horizontal plane (azimuth) due to obstacles, e.g., buildings. Usually, the elevation angle has a small impact on the signal propagation, as the height of buildings and the RIS panel placement typically result in a fixed elevation angle, particularly for ground-level UEs.} For each UE position, defined by the steering direction $\phi_{\rm UE}$, we generate the phase-shift matrix $\mathbf{\Psi}$, where the entries are determined using the following equation \cite{EBasar2019}:
	\begin{align}
		\psi_{n,m}=&\exp\left(-j(\angle h_{{\rm BS}, n,m}+\angle h_{n,m, {\rm UE}})\right), \nonumber\\&\qquad \quad\forall n\in\{1,\cdots, N_{\rm z}\}~\&~\forall m\in\{1,\cdots, N_{\rm y}\}.
		\label{phase-shift}
	\end{align}
	The objective is to cover the region defined by the azimuth angle $\phi_{\rm UE}$ spanning from $-80^{\circ}$ to $80^{\circ}$, and generate the phase-shift matrix for each steering direction $\phi_{\rm UE}$. To achieve this, we follow the procedure described in Algorithm 1. The process is repeated for each of the three deployment cases described in section III-B.
	
	\begin{algorithm}[h!]
		\caption{RIS Phase-Shift Design and Beam Steering}
		\begin{algorithmic}[1]
			\REQUIRE Number of RIS elements $N$, BS angles $(\phi_{\rm BS}, \theta_{\rm BS})$, UE angle $\theta_{\rm UE}$, distances $d_{\rm BS,RIS}$ and $d_{\rm RIS,UE}$
			
			\STATE Initialize the steering angle: $\phi_{\rm UE} = -80^\circ$
			\STATE Generate channel dataset using equations~(\ref{Chan1}) and (\ref{Chan2})
			\STATE Compute initial phase-shift matrix $\mathbf{\Psi}_1$ using~(\ref{phase-shift}) at $\phi_{\rm UE}$
			\STATE Evaluate RIS gain $G_{\rm RIS}(\phi)$ for all $\phi \in [-80^\circ\!:\!1^\circ\!:\!80^\circ]$
			\STATE Find angle $\phi_{\min}$ such that $G_{\rm RIS}(\phi_{\min}) = G_{\rm RIS,min}$
			
			\STATE Set iteration index $q = 1$
			\WHILE{$\phi_q \leq 80^\circ$}
			\STATE Generate channel dataset using~(\ref{Chan1}) and (\ref{Chan2})
			\STATE Compute phase-shift matrix $\mathbf{\Psi}_q$ using~(\ref{phase-shift}) at $\phi_q$
			\STATE Evaluate RIS gain $G_{\rm RIS}(\phi)$ for all $\phi \in [-80^\circ\!:\!1^\circ\!:\!80^\circ]$
			\STATE Identify $\phi_q$ where $G_{\rm RIS}(\phi_q) = G_{\rm RIS,min}$
			\STATE Compute phase difference: $\Delta \phi = \phi_q - \phi_{q-1}$
			\STATE Update steering angle: $\phi_{q+1} = \phi_q + \frac{\Delta \phi}{2}$
			\STATE Compute next phase-shift matrix $\mathbf{\Psi}_{q+1}$
			\STATE Increment index: $q \gets q + 1$
			\ENDWHILE
			\ENSURE $Q=q-1$, Phase-shift matrices $\{\mathbf{\Psi}_q\}_{q=1}^{Q}$
		\end{algorithmic}
	\end{algorithm}

	\begin{figure}[h!]
		\centering
		\hspace*{-0.2in}
		\centering
		\includegraphics[height=3.2in, width=4in]{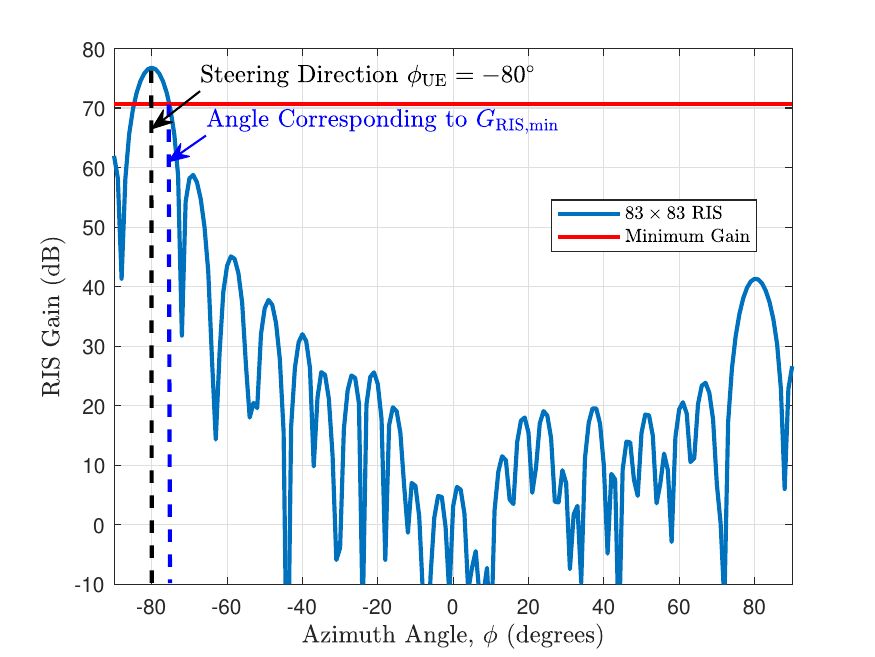}
		\caption{RIS gain versus azimuth angle $\phi$ for a steering direction $\phi_{\rm UE}=-80^{\circ}$ in {\it Deployment Case 1}, i.e., RIS near the BS. Here, $\Delta G^{\rm dB}=6$ dB.}
		\label{fig:prob1_6_21}
	\end{figure}
	
	The plots in Fig.~\ref{fig:prob1_6_21} illustrate the RIS gain as a function of the azimuth angle $\phi \in [-80^{\circ},80^{\circ}]$ for a steering direction $\phi_{\rm UE} = -80^{\circ}$, along with the corresponding minimum gain in this configuration. It is observed that the RIS gain attains its minimum value, namely, $G_{\rm RIS,min} = 70.625$ dB, at approximately $\phi_q = -76^{\circ}$. This angle is then used to compute the subsequent steering direction, as outlined in Algorithm 1. The same procedure is iteratively applied to determine the successive angle $\phi_{q+1}$ and the associated phase-shift matrix $\mathbf{\Psi}_{q+1}$, as described in Algorithm 1, continuing until the steering direction reaches $\phi_{\rm UE}=80^{\circ}$.
	\subsection{Correlation Analysis}
	We now outline the method to design the new RIS based on correlation. The process for identifying correlated elements consists in the following steps:
	\begin{itemize}
		\item {\bf Selecting Element Pairs:} Computing the phase-shift differences between all possible pairs of entries in matrix $\mathbf{\Psi}_q$, and storing the results in the vector $\Delta \mathbf{\Psi}_q$. This process is repeated for all phase-shift matrices $\{\mathbf{\Psi}_q\}_{q=1}^{Q}$ generated using Algorithm 1, and the corresponding differences are stored in the vectors $\{\Delta \mathbf{\Psi}_q\}_{q=1}^{Q}$.
		\item {\bf Defining the Correlation Threshold:} A predefined threshold, $\psi_{\rm th}$, determines whether two elements are correlated. The threshold is set based on the desired correlation level and the specific requirements of the deployment scenario.
		\item {\bf Quantifying Correlation:} To capture the correlation state of each element, we use a binary variable, $b_m \in \{0,1\}$ for $m \in\{1, 2, \ldots, M\}$, where $M = \binom{N^2}{2}$ is the total number of element pairs. If the phase-shift difference between a pair of elements $\Delta \Psi_{m, q}$, $q\in\{1, \cdots, Q_{\rm sc}\}$, is less or equal to the predefined threshold across all the scenarios, the elements are considered as correlated and $b_m$ is set to unity; otherwise $b_m=0$.
	\end{itemize}
	
	The binary values are stored in a vector $\mathbf{b} = [b_1, b_2, \ldots, b_M]$, which represents the correlation among the $M$ pairs of phase shifting elements. If the correlation vector contains only ones, the RIS is deemed as fully connected. When $b_m = 1$, the two elements corresponding to the $m^{\rm th}$ combination are correlated and can be connected together.
	
	Algorithm \ref{alg:compute-diff} summarizes the correlation analysis process, detailing the steps for generating the phase-shift matrices, computing the phase differences, and identifying the correlated elements. The correlation process is conducted for each of the three deployment cases described in section III-B.
	
	\begin{algorithm} [h!]
		\caption{Correlation Analysis}\label{alg:compute-diff}
		\begin{algorithmic}[1]
			\REQUIRE $\mathbf{\Psi}_{q},~\forall q\in\{1, \cdots, Q\}$; $\psi_{\rm th}$.\\
			\STATE Initialize $\mathbf{b}=\mathbf{0}_{M\times 1}$, $\Delta \mathbf{\Psi}_q=\mathbf{0}_{M\times 1}$, and $\mathbf{K}_{\rm pairs}=\mathbf{0}_{M\times 4}$, for storing the binary decisions, phase-shift differences and index pairs, respectively.\\ 
			\STATE For $q\in\{1, \cdots, Q\}$, loop over all elements pairs in matrix $\mathbf{\Psi}_q$ to compute their differences, and save the values in $\Delta \mathbf{\Psi}_q$.\\
			\STATE Iterate over the vectors $\Delta \mathbf{\Psi}_q$, $q\in\{1, \cdots, Q\}$, comparing the differences element-wise to check if they are within the specified threshold.\\
			\quad {\bf If} $\Delta \Psi_{m,q} \leq \psi_{\rm th}$ for $q\in\{1, \cdots, Q\}$\\
			\quad Set $b_m=1$.\\
			\quad Extract the corresponding indexes of the elements.\\
			\quad Store the indexes into matrix $\mathbf{K}_{\rm pairs}$.\\
			\quad {\bf Else}\\
			\quad Set $b_m=0$.\\
			\quad {\bf End If}
			\STATE Store $\mathbf{I}_{\rm x}$ and $\mathbf{b}$ in matrix $\mathbf{F}=[\mathbf{K}_{\rm pairs} \mathbf{b}]$.\\
			\ENSURE Matrix $\mathbf{F}$ containing the index pairs and binary decisions.
		\end{algorithmic}
	\end{algorithm}	
	
	Since the threshold value $\psi_{\rm th}$ directly impacts the number of correlated elements, next we analyze this relationship.
	
	\begin{figure}[h!]
		\centering
		\hspace*{-0.2in}
		\includegraphics[height=3.2in, width=4in]{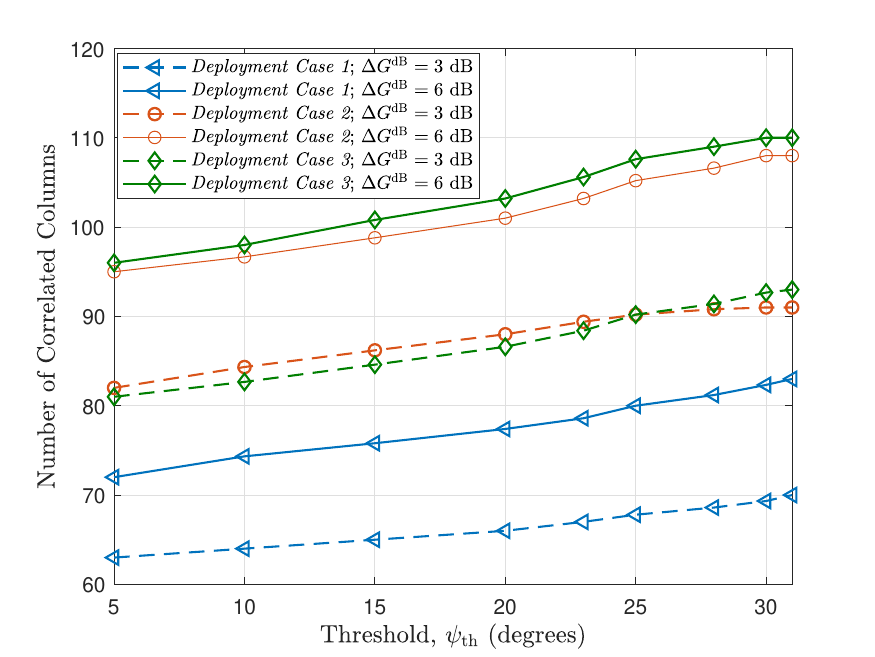}
		\caption{Number of correlated columns versus the threshold value $\psi_{\rm th}$.}
		\label{dcc256}
	\end{figure}
	
	Figure \ref{dcc256} illustrates the relationship between the number of correlated columns and the threshold value $\psi_{\rm th}$ for the three deployment cases described in section III-B. For instance, in deployment case 1, where the RIS is near the BS, the maximum number of correlated elements occurs at $\psi_{\rm th} = 31^{\circ}$ for $\Delta G^{\rm dB} = 3$ and at $\psi_{\rm th} = 30^{\circ}$ for $\Delta G^{\rm dB} = 6$, indicating optimal threshold values for these configurations. At lower threshold values, such as $\psi_{\rm th} = 15^{\circ}$, all scenarios exhibit a notably reduced number of correlated columns. This suggests that increasing $\psi_{\rm th}$ enhances the correlation among the columns. Furthermore, the variation in the correlation trends across the three cases underscores the significant impact of the deployment configuration on the nature of the correlation.
	
	\begin{figure*}[t]
		\centering
		\includegraphics[height=4in, width=6in]{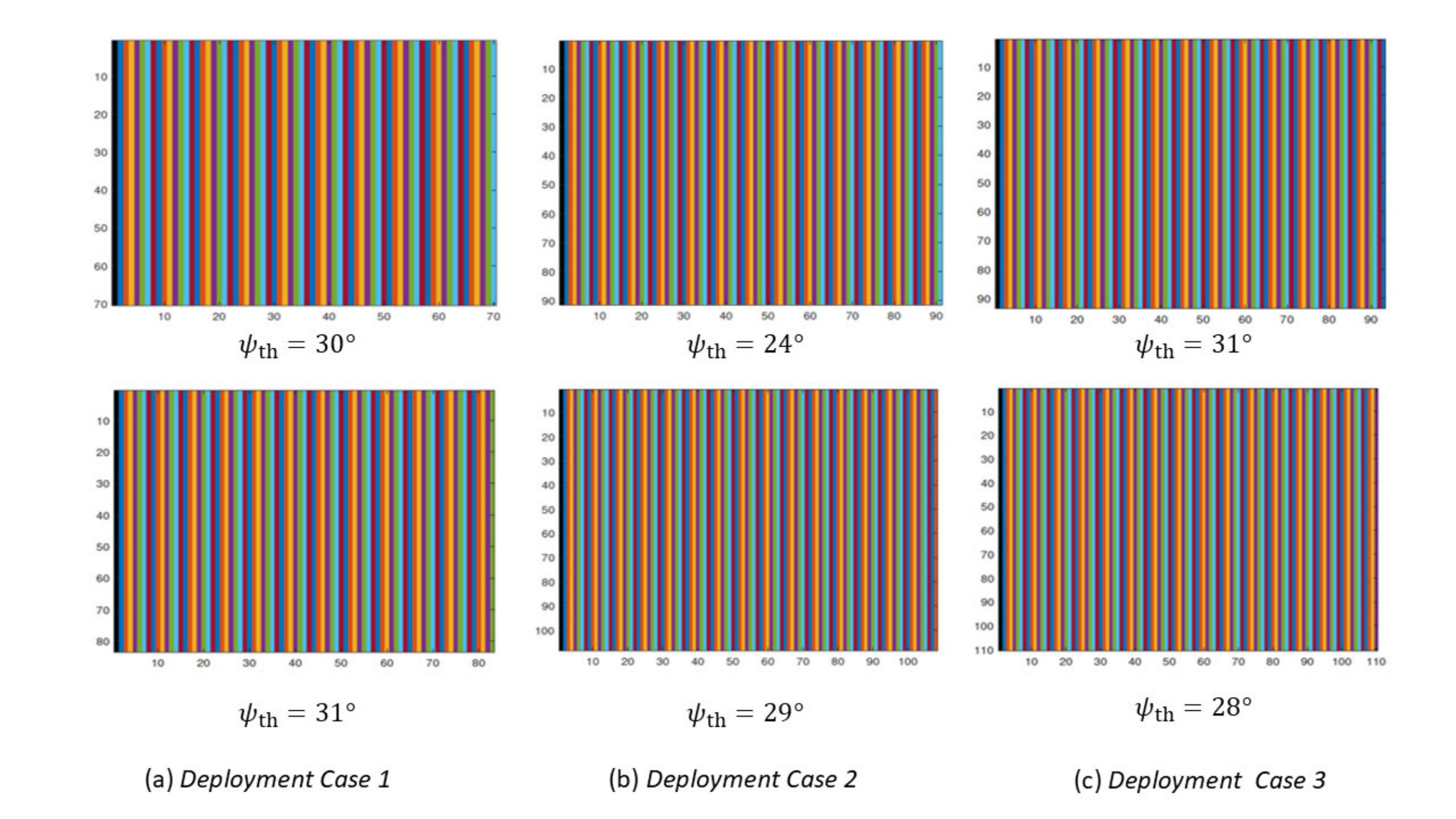}
		\caption{Illustration of correlated elements in the three deployment cases: RIS elements sharing the same color maintain the same phase-shift difference as the steering direction, $\phi_{\rm UE}$ varies from $-80^{\circ}$ to $80^{\circ}$. The figures in the first row correspond to $\Delta G^{\rm dB}=3$ dB, while the figures in the second row correspond to $\Delta G^{\rm dB}=6$ dB.}
		\label{fig67}
	\end{figure*}
	
	Figure~\ref{fig67} shows the results obtained by applying the proposed methods (Algorithm 1 and Algorithm 2) for the three deployment cases under consideration. The threshold values used are respectively $\psi_{\rm th} = 30^{\circ}$ and $31^{\circ}$, $24^{\circ}$ and $29^{\circ}$, $31^{\circ}$ and $28^{\circ}$. The figure demonstrates a strong correlation along the columns of the RIS, suggesting that the elements within each column can be grouped together and controlled using a single control signal. This implies that, for effective beamforming in the desired direction, it is sufficient to configure the phase shifts of the elements in the first row of the RIS. This leads to the proposed RIS structure, namely, the Connected-RIS. This structure ensures that the imposed phase pattern propagates consistently across the remaining elements in each column of the surface.	
	\section{Performance Evaluation of the Connected-RIS}	
	\subsection{Control and Load Impedance}
	In the Connected-RIS architecture, the first element of each column is a reconfigurable unit modeled by a load impedance $Z_i$, where $i \in \{ 1, \dots, N_{\rm z}\}$. Here, $N_{\rm z}$ denotes the number of groups of correlated elements, which corresponds to the number of columns in the structure. For comparison purposes, two benchmark designs are considered, both based on fully controlled elements: (i) the min-gain RIS, which corresponds to the RIS structure with the minimum number of elements that satisfies the fair coverage criterion (cf. section III); and (ii) the $\Delta G^{\rm dB}$-RIS, which refers to the conventional RIS where the number of elements is selected to achieve the required gain after adding a gain margin $\Delta G^{\rm dB}$ = 3dB or 6dB.
	
	\begin{figure}[h!]
		\hspace*{-0.15in}
		\centering
		\includegraphics[height=3in, width=3.7in]{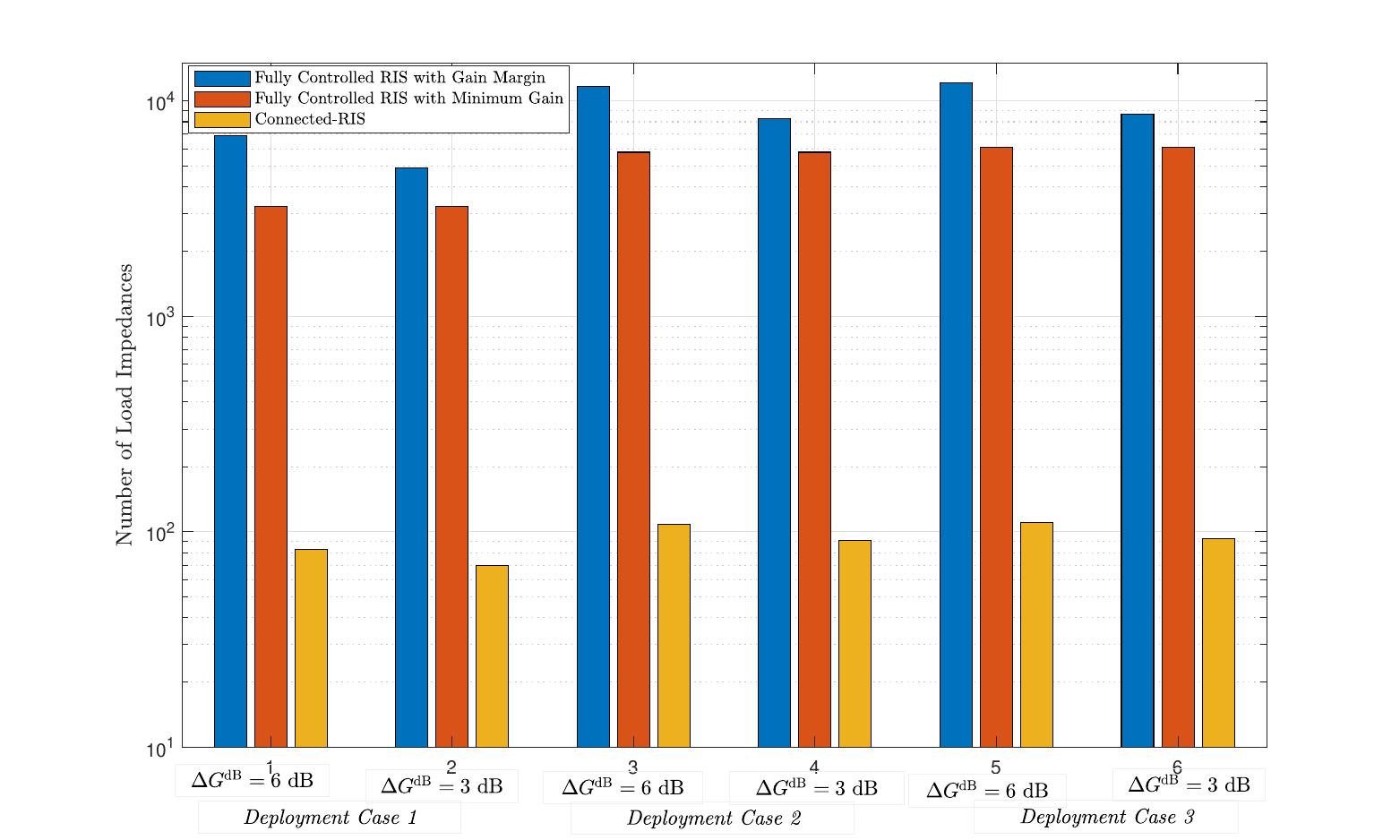}
		\caption{Number of load impedances: comparison of Connected-RIS versus two benchmarks in different deployment cases.}
		\label{fig:prob1_2}
	\end{figure}
	
	Figure~\ref{fig:prob1_2} illustrates that the Connected-RIS architecture significantly reduces the number of required load impedances compared to the fully controlled RIS structures (3dB-RIS or 6dB-RIS, and the min-gain RIS) across all the deployment cases. This reduction is consistently observed for both margins, i.e., $\Delta G^{\rm dB}=6$ dB and $\Delta G^{\rm dB}=3$ dB, with the Connected-RIS exhibiting an exponential decrease in impedance requirements, as evidenced by the logarithmic scale. This substantial reduction leads to lower hardware complexity, reduced implementation costs, and a more simplified design of the DC control circuitry, thereby positioning the Connected-RIS as a more efficient and scalable architecture.
	
	For the proposed Connected-RIS design, we introduce a DC control circuit, where each load impedance is driven by a single DC bias control line, with a shared ground connection used as the return path. In total, the Connected-RIS requires 83 DC control lines, compared to 6889 control lines for the fully controlled 6dB-RIS, and 70 DC biasing instead of 4900 control lines required for the fully controlled 3dB-RIS. In both configurations, the Connected-RIS also outperforms the fully controlled with minimum gain RIS, which requires 3249 control lines. This substantial reduction in the number of control lines highlights the markedly lower complexity of the DC control circuit in the Connected-RIS design, making it a more practical and scalable alternative to the fully controlled RIS architectures.
	
	\begin{figure}[h!]
		\centering
		\hspace*{-0.45in}
		\includegraphics[height=3in, width=3.7in]{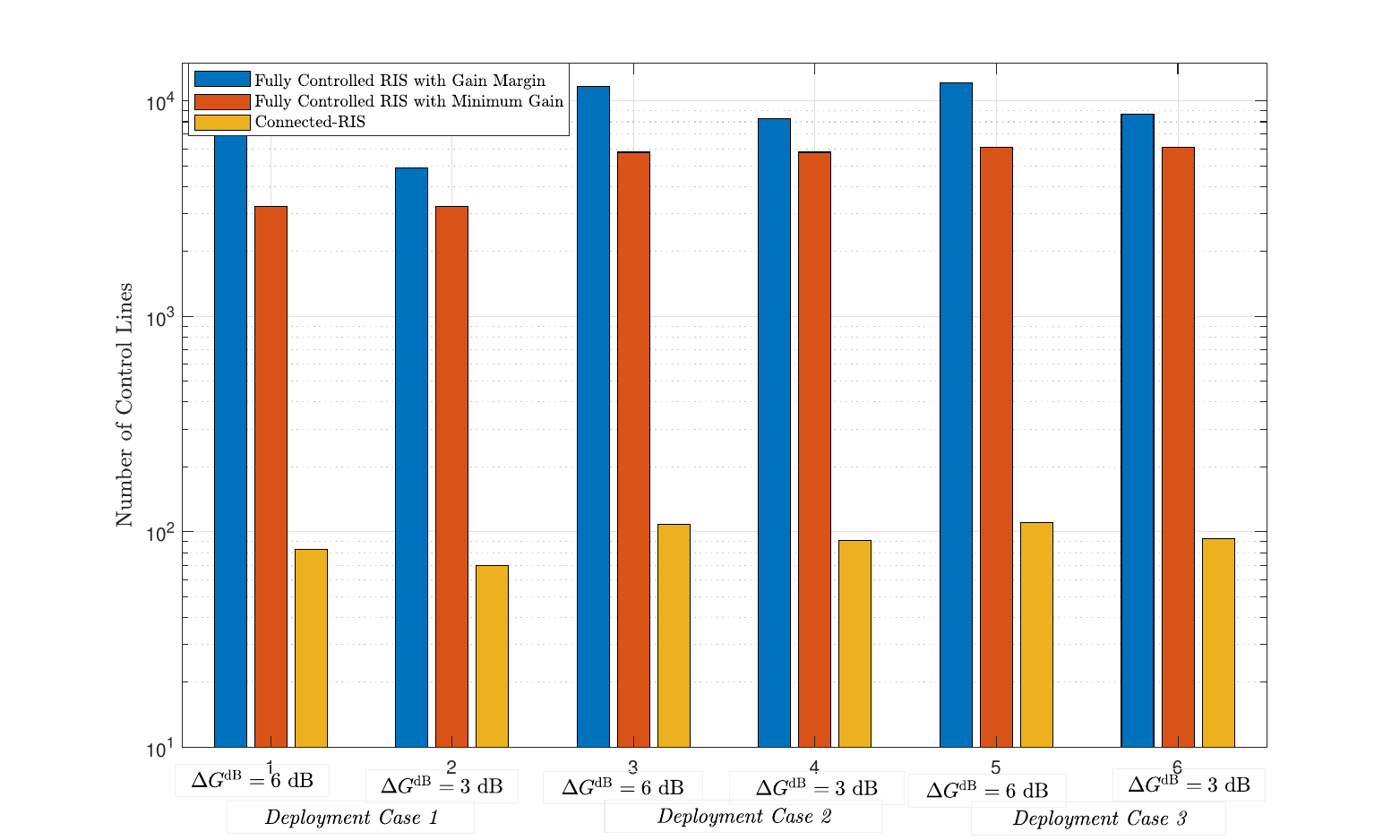}
		\caption{Number of control lines: comparison of Connected-RIS versus two benchmarks in different deployment cases.}
		\label{fig:prob1_1}
	\end{figure}
	
	A further comparison in terms of the number of BS-RIS control signals can also be considered. We find that the number of control signals transmitted from the BS to the RIS via the separated control link is $N^2$ for the fully controlled RIS (3dB-RIS, 6dB-RIS and min-gain RIS), where $N$ is the total number of RIS elements ($N=N_{\rm req}$ or $N=N_{\rm min}$ according to the structure); and $N_{\rm z}^2$ for the Connected-RIS, with $N_{\rm z}$ is the number of correlated elements (columns), which is always less than that of the fully controlled RIS structures.
	\subsection{Power Consumption Analysis}
	Now, we evaluate the power consumption of the proposed architecture. The power consumption of a RIS can be divided into two main components, the static power and the unit cell power \cite{wang2023static}. The static power, denoted $P_{\rm static}$, corresponds to the power required by the control board and drives the circuits, e.g., registers. Power $P_{\rm units}$ represents the power consumed by the RIS unit cells, which varies depending on the type of switching elements used. Based on this description, the power consumption of the RIS can be expressed as
	\begin{align}
		P_{\rm {\small \sum}}&=P_{\rm static}+P_{\rm units},
	\end{align}
	\noindent where $P_{\rm static}=P_{\rm control}+P_{\rm circuit}$, in which $P_{\rm control}$ is the power consumption of the control board, which can be considered as constant, and $P_{\rm circuit}$ is the power consumption of the drive circuits. The latter depends on the type of adjustable electronic components, the number of control signals, and the self-power consumption characteristics \cite{wang2024}. Based on that, $P_{\rm circuit}$ of the Connected-RIS can be formulated as 
	\begin{align}
		P_{\rm circuit}=\left\lceil\frac{N_{\rm c}}{N_{\rm z} N_{\rm s}}\right\rceil P_{\rm drive},
	\end{align}
	\noindent where $N_{\rm c}$ is the number of adjustable electronic components, $N_{\rm s}$ is the number of control signals, $N_{\rm z}$ is the number of correlated elements, which corresponds to the number of columns of the RIS, $\left\lceil \cdot \right\rceil$ is the ceiling function, and $P_{\rm drive}$ is the power consumption of the drive circuit, which depends on the bit resolution of the reflecting elements.
	
	For comparison purposes, we also provide the expression for the circuit power consumption corresponding to the the fully controlled RIS (with gain margin and with minimum gain), which is given by
	\begin{equation}
		P_{\rm circuit}=\left\lceil\frac{N_{\rm c}}{N N_{\rm s}}\right\rceil P_{\rm drive}.
	\end{equation}
	
	Let $P_{\rm unit}$ denote the power consumption of a reflecting element along with the supporting biasing circuit (unit cell). In this case, the reflecting element consists of three PIN diodes, with the PIN diode individual power denoted by $P_{\rm diode}$. Therefore, the total power of each unit cell is given by $P_{\rm unit}=3P_{\rm diode}$. Hence, the total power of the RIS units is obtained as $P_{\rm units}=N_{\rm units}P_{\rm unit}$, where $N_{\rm units}$ is the total number of elements.  In this work, we define the number of units as either $N_{\rm units}=N$, for the fully controlled RIS or $N_{\rm units}=N_{\rm z}=N_{\rm y}$ for the Connected-RIS.
	
	The FPGA XC7K70T is embedded as the control board, with a working voltage of 24 V and a current of 0.2 A \cite{wang2023static}. Hence, $P_{\rm control}=P_{\rm FPGA}=4.8$ W. We assume that the power consumption of a RIS unit is $P_{\rm unit}=15$ mW. Here, we employ a 4-bit shift register, which typically consists of three D flip-flops connected in series to control each element (TTL 74LS194), thus $P_{\rm drive}=75$ mW. Based on these values, the total power consumption of a RIS panel in the three design configurations, namely, the fully controlled RIS with gain margin (the 3dB-RIS, the 6dB-RIS), the fully controlled RIS with minimum gain (the min-gain RIS), and the Connected-RIS, is compared in Table \ref{tab:merged_mos} for the three deployment cases described in section III-B.
	
	\begin{table*}[h!]
		\centering
		\caption{RIS Power Consumption for Different Deployment Cases.}
		\label{tab:merged_mos}
		\begin{tabular}{|p{2.5cm}||p{2.7cm}|p{2.7cm}|p{1.9cm}|p{1.9cm}|p{2.5cm}|}
			\hline
			\multicolumn{6}{|c|}{\bf RIS Power Consumption (W)} \\
			\hline
			{\bf Deployment Cases} & Connected-RIS (3 dB) & Connected-RIS (6 dB)& 3dB-RIS & 6dB-RIS & Min-Gain RIS\\
			\hline
			\textit{Case 1} & 5.860 & 6.045 & 78.410 & 108.240  &43.815 \\
			\hline
			\textit{Case 2} & 6.170 & 6.420 & 129.127  & 179.87 &91.440 \\
			\hline
			\textit{Case 3} & 6.200  & 6.450 & 134.647  & 186.412 & 96.060 \\
			\hline
		\end{tabular}
	\end{table*}
	
	A key finding from Table \ref{tab:merged_mos} is that the Connected-RIS structure, designed based on the correlation analysis, offers substantial power savings compared to the fully controlled structures (3dB-RIS, 6dB-RIS and the min-gain RIS). For example, in {\it Deployment Case 1}, the Connected-RIS consumes only 5.86 W, whereas the 3dB-RIS and the min-gain RIS consume 78.41 W and 43.82 W, respectively. This corresponds to a power reduction of approximately 92.5\% relative to the 3dB-RIS architecture, and 86.6\% relative to the min-gain one. Moreover, the Connected-RIS exhibits robustness to gain margin variations, with only a minor increase in the power consumption when the gain margin is increased from 3 dB to 6 dB, unlike the fully controlled RIS, which experiences a significant rise. For all deployment cases, the proposed RIS architecture consistently maintains low power consumption, even as the distance of the RIS from the BS increases, demonstrating its scalability and adaptability.
	\subsection{Gain Analysis}
	Now, we conduct a performance evaluation of the RIS gain achieved by the Connected-RIS design, in comparison with the fully controlled designs.
	
	\begin{figure}[h]
		\centering
		\begin{minipage}[t]{0.2\linewidth}
			\hspace*{-1.05in}	\includegraphics[width=1.7in,height=2.1in]{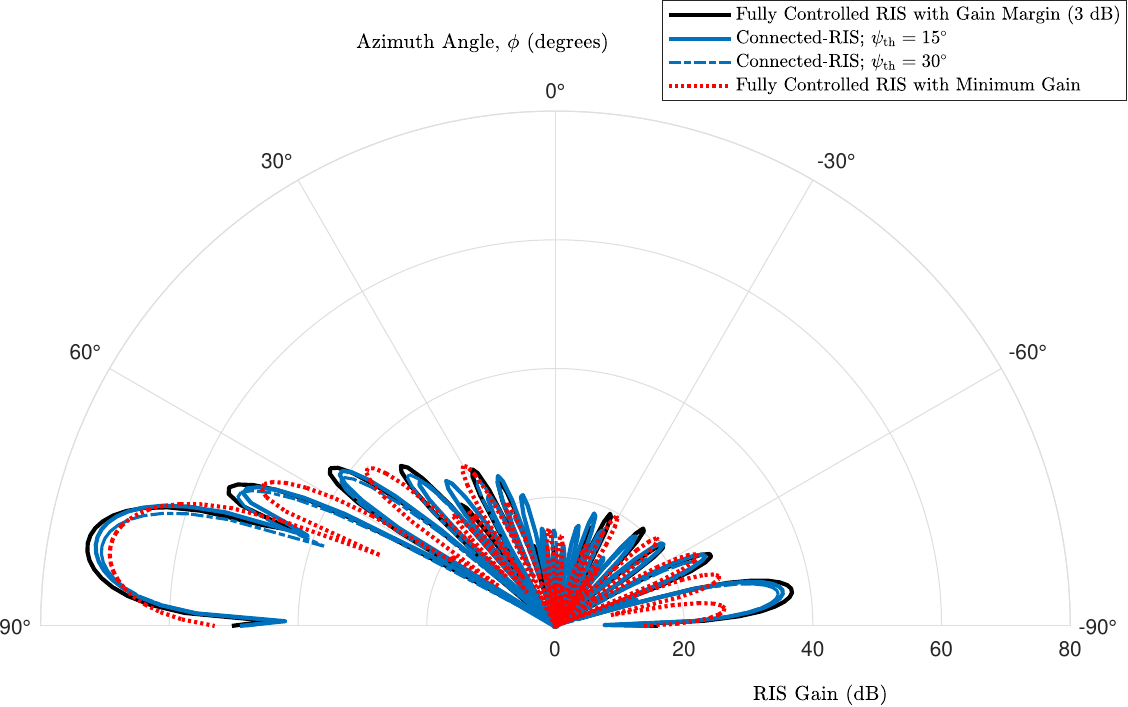}
		\end{minipage}
		\begin{minipage}[t]{0.2\linewidth}
			\centering
			\hspace*{-0.05in}	\includegraphics[width=1.7in,height=2.1in]{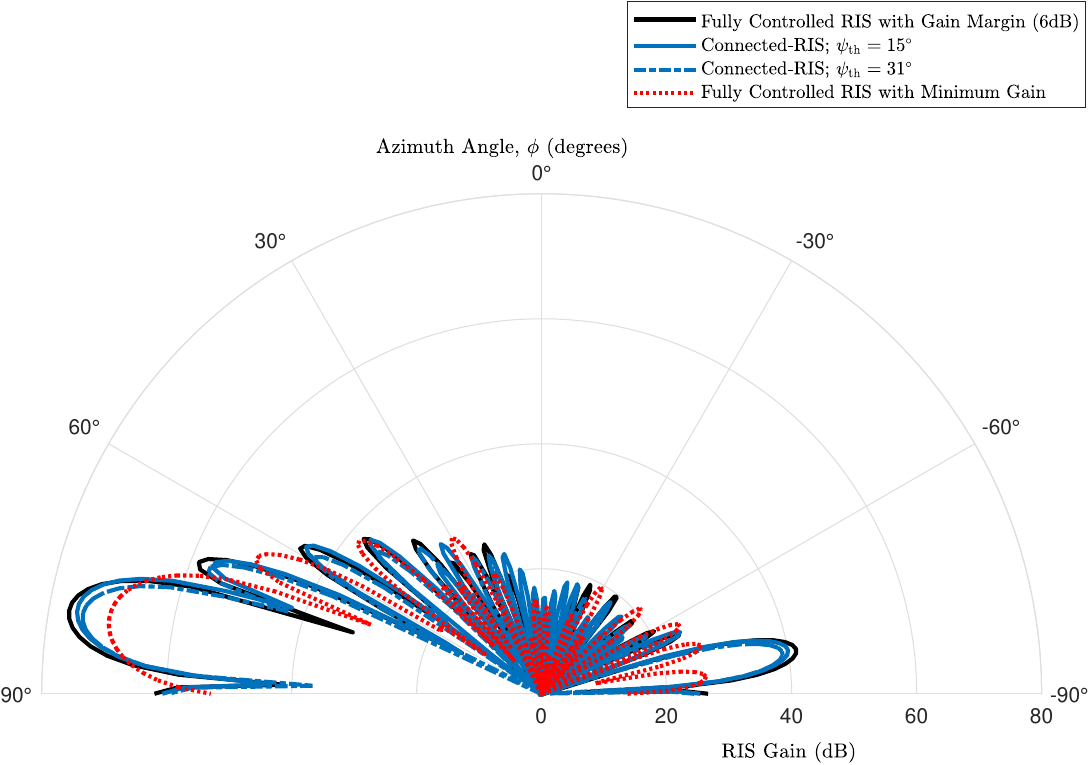}
		\end{minipage}
		\caption{RIS gain versus azimuth angle $\phi$ in {\it Deployment Case 1}, with $\phi_{\rm UE}=80^{\circ}$: (a) $\Delta G^{\rm dB}=3$ dB, (b) $\Delta G^{\rm dB}=6$ dB.}
		\label{fig:prob1_6_6}
	\end{figure}
	
	\begin{figure}[h]
		\centering
		\begin{minipage}[t]{0.2\linewidth}
			\hspace*{-1.05in}
			\includegraphics[width=1.7in,height=2.1in]{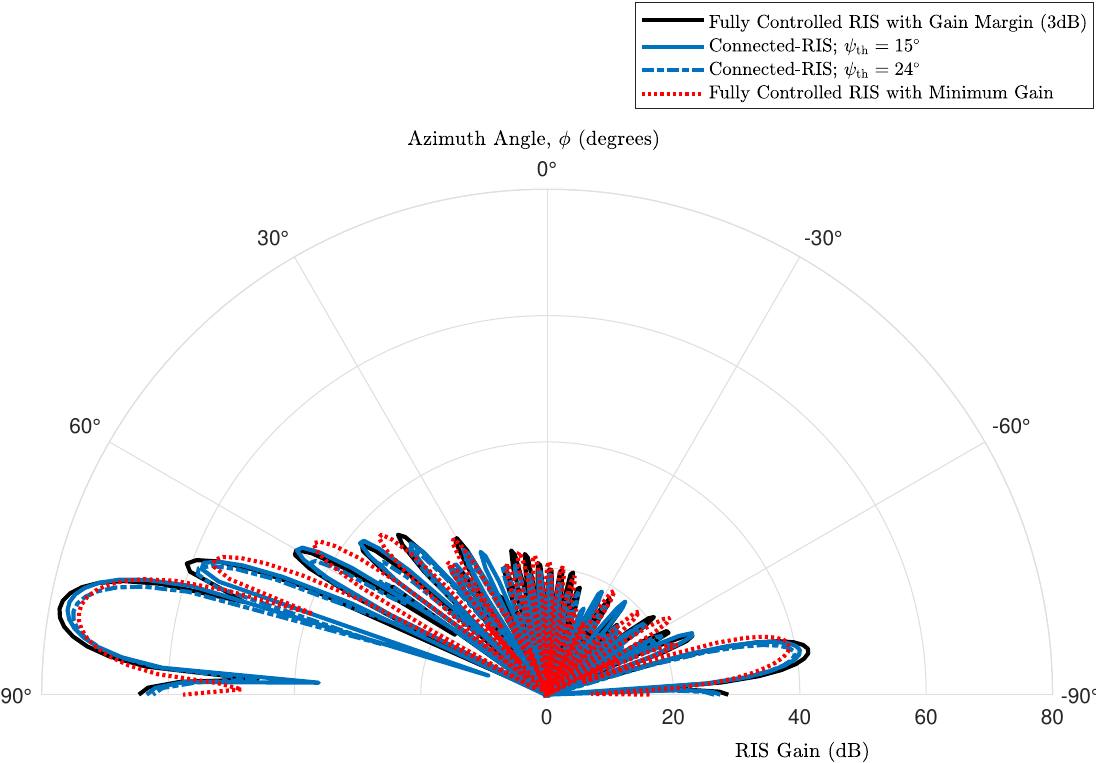}
		\end{minipage}
		\begin{minipage}[t]{0.2\linewidth}
			\centering
			\hspace*{-0.05in}
			\includegraphics[width=1.7in,height=2.1in]{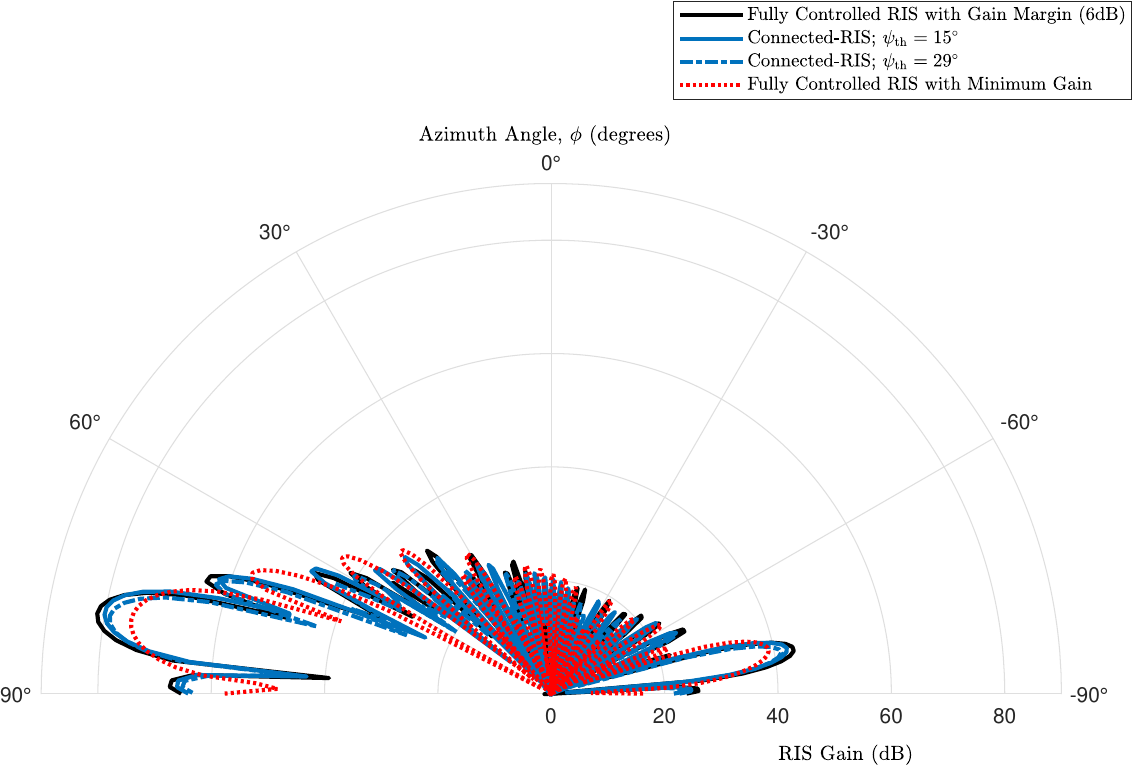}
		\end{minipage}
		\caption{RIS gain versus azimuth angle $\phi$ in {\it Deployment Case 2}, with $\phi_{\rm UE}=80^{\circ}$: (a) $\Delta G^{\rm dB}=3$ dB, (b) $\Delta G^{\rm dB}=6$ dB.}
		\label{fig:prob1_6_71}
	\end{figure}
	
	\begin{figure}[h]
		\centering
		\begin{minipage}[t]{0.2\linewidth}
			\hspace*{-1.05in}
			\includegraphics[width=1.7in,height=2.1in]{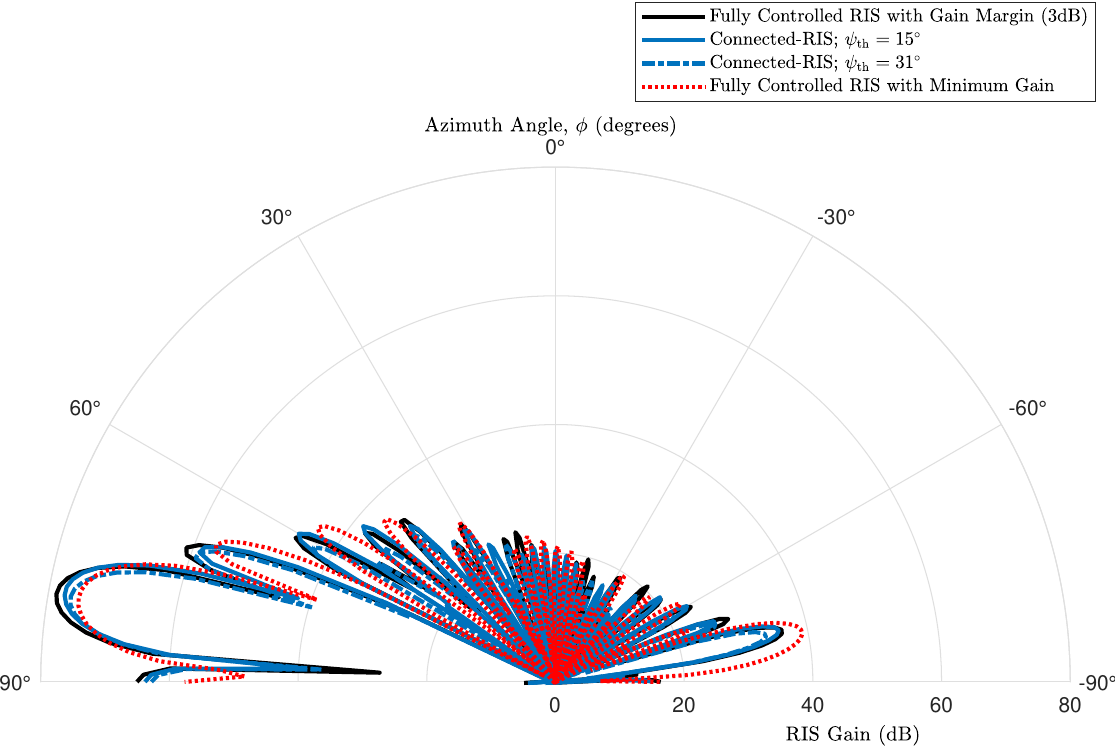}
		\end{minipage}
		\begin{minipage}[t]{0.2\linewidth}
			\centering
			\hspace*{-0.05in}
			\includegraphics[width=1.7in,height=2.1in]{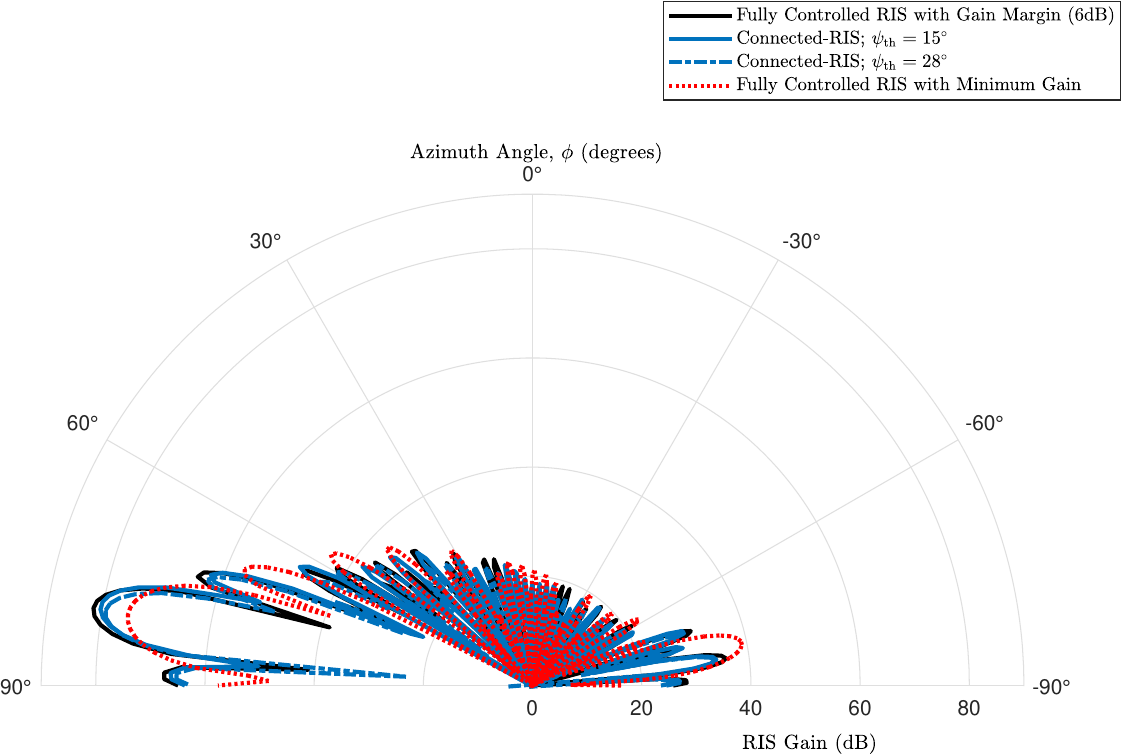}
		\end{minipage}
		\caption{RIS gain versus azimuth angle $\phi$ in {\it Deployment Case 3}, with $\phi_{\rm UE}=80^{\circ}$: (a) $\Delta G^{\rm dB}=3$ dB, (b) $\Delta G^{\rm dB}=6$ dB.}
		\label{fig:prob1_6_72}
	\end{figure}

	Figures~\ref{fig:prob1_6_6}–\ref{fig:prob1_6_72} present a comprehensive evaluation of the RIS gain performance across various configurations: the fully controlled RIS with gain margin, the fully controlled RIS with minimum gain, and the Connected-RIS, in the deployment cases described in section III-B, considering different threshold values $\psi_{\rm th}$. For instance, as shown in Fig.~ \ref{fig:prob1_6_71}, the fully controlled and the Connected-RIS exhibit similar gain characteristics, with minor angular shifts in the Connected-RIS nulls. Although the maximum gain of the Connected-RIS remains slightly below that of the fully controlled RIS with gain margin, it continues to exceed the performance of the fully controlled RIS with minimum gain, thereby ensuring compliance with fair coverage requirements. While the fully controlled with gain margin offers high performance, its excessive power consumption renders it unsuitable for large-scale deployments (cf. Table II). Although the fully controlled RIS with minimum gain is more efficient than that with gain margin ($\Delta G^{\rm dB}$-RIS) configurations, it remains less energy-efficient than the Connected-RIS. These results underscore the Connected-RIS as a compelling architecture for next-generation wireless networks, offering a favorable trade-off between performance and energy efficiency, particularly in energy-constrained and large-scale scenarios. Overall, the Connected-RIS demonstrates a strong ability to balance the control complexity with the system performance, and to reliably meet the fair coverage condition in diverse deployment conditions.	
	\subsection{Rate Analysis}
	While reducing power consumption is essential for enhancing the energy efficiency, it is equally important to ensure that this improvement does not compromise the communication rates or the overall system performance. Therefore, evaluating the achievable rate is crucial for determining whether the Connected-RIS design maintains an acceptable trade-off between the energy efficiency and the communication effectiveness. In these investigations, we set $N_0=-174+10\log_{10}(B)$ dBm, where $B = 1$ MHz is the system bandwidth \cite{Li2023}.
	
	\begin{figure}[h]
		\centering
		\hspace*{-0.2in}	\includegraphics[width=4in,height=3.2in]{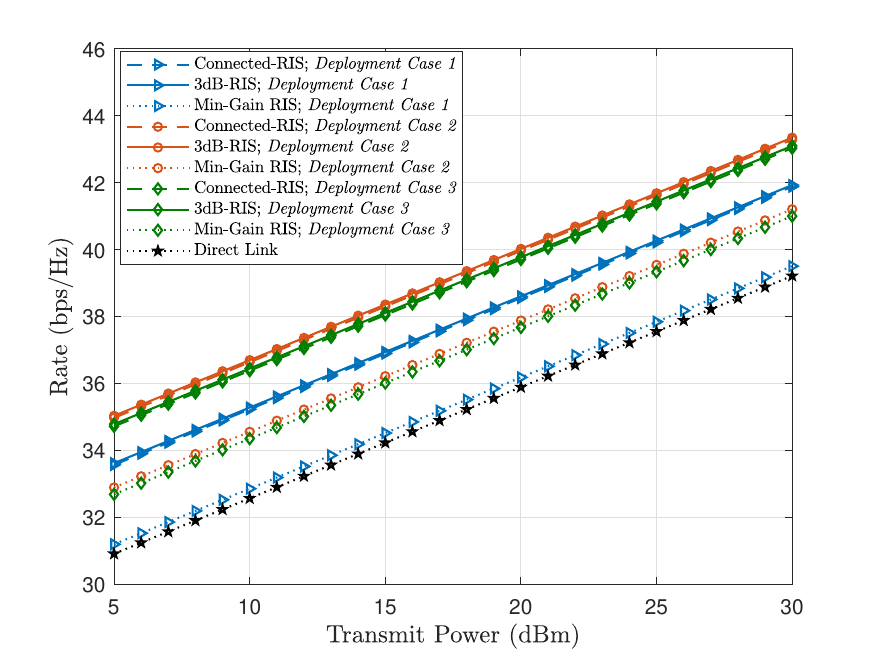}
		\caption{Rate versus transmit power for $\Delta G^{\rm dB}=3$ dB and $\phi_{\rm UE}=80^{\circ}$.}
		\label{figconfig21-a}
	\end{figure}
	
	\begin{figure}[h]
		\centering
		\hspace*{-0.2in}	\includegraphics[width=4in,height=3.2in]{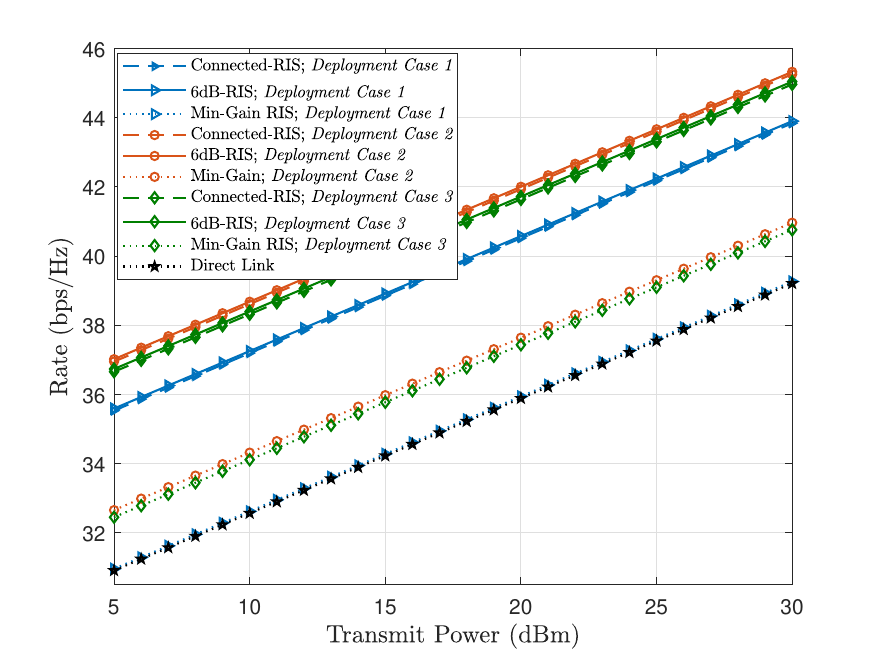}
		\caption{Rate versus transmit power for $\Delta G^{\rm dB}=6$ dB and $\phi_{\rm UE}=80^{\circ}$.}
		\label{figconfig21-b}
	\end{figure}
	
	In Figs.~\ref{figconfig21-a} and ~\ref{figconfig21-b}, the achieved rate at the typical UE is plotted for the Connected RIS, and the fully controlled RIS configurations. The results reveal that all configurations attain nearly identical rates. This outcome can be attributed to the fact that the three RIS structures are designed to optimize the phase shifts of the incident signals, thereby enhancing the received signal strength at the UE. Furthermore, since all the configurations are deployed with the same resource allocation strategies, their ensuing performance remains comparable despite differences in their architectural complexity. These findings underscore a key advantage of the Connected-RIS: it offers comparable performance to the fully controlled RIS, while significantly reducing the power consumption and control overhead. As a result, the Connected-RIS emerges as a more energy-efficient and cost-effective alternative that still satisfies the quality-of-service requirements as the conventional designs.
	\subsection{RIS Configuration and Case Study}
	It is important to recall that the objective is to ensure fair coverage over the angular region defined by the azimuth range $\phi=-80^{\circ}$ to $\phi=80^{\circ}$ at a fixed elevation angle $\theta_{\rm RIS}$, while maintaining constant azimuth angle $\phi_{\rm BS}$ and elevation angle $\theta_{\rm BS}$ at the BS. To achieve the maximum gain throughout the coverage area, the RIS must be continuously reconfigured to steer its beam towards the azimuth direction $\phi$, such that the resulting gain $G_{\rm RIS}(\phi)$ meets or exceeds a predefined threshold value $G_{\rm th}$. The minimum number of distinct codewords required to achieve full coverage under this criterion can be systematically determined using Algorithm 3.
	
	\begin{algorithm}[h!]
		\caption{Computation of the Required Number of Codewords}
		\begin{algorithmic}[1]
			\REQUIRE Gain threshold $G_{\rm th}$.
			\STATE Initialize the number of configurations: $N_{\rm conf} \leftarrow 0$.
			\STATE Set initial steering angle: $\phi_{\rm UE} \leftarrow -80^\circ$.
			\STATE Compute RIS gain $G_{\rm RIS}(\phi)$ for all $\phi \in [-80^\circ, 80^\circ]$ with $1^\circ$ resolution.
			\STATE Find the first angle $\phi_1$ such that $G_{\rm RIS}(\phi_1) = G_{\rm th}$.
			\STATE Set $q \leftarrow 2$.
			\WHILE{$\phi_q \leq 80^\circ$}
			\STATE Compute RIS gain $G_{\rm RIS}(\phi)$ for all $\phi \in [-80^\circ, 80^\circ]$.
			\STATE Find the next angle $\phi_q$ such that $G_{\rm RIS}(\phi_q) = G_{\rm th}$.
			\STATE Update configuration count: $N_{\rm conf} \leftarrow N_{\rm conf} + 1$.
			\STATE Increment index: $q \leftarrow q + 1$.
			\ENDWHILE
			\ENSURE Required number of configurations $N_{\rm conf}$.
		\end{algorithmic}
	\end{algorithm}

	\subsubsection{Configuration Approaches}
	By applying Algorithm 3, the number of codewords required to achieve full coverage is determined when using the Connected-RIS or the fully controlled RIS designs. For instance, to span the angular range from $-80^{\circ}$ to $80^{\circ}$, the $83 \times 83$ Connected-RIS requires only 52 codewords, whereas the min-gain RIS necessitates 80 codewords. Based on this analysis, two deployment strategies can be considered. In the first scenario, the RIS is dynamically reconfigured using a register that stores all predefined codewords, allowing the beam to be steered in real time towards the desired direction. In the second scenario, 52 Connected-RIS panels of size $83 \times 83$ are deployed, each pre-configured with a fixed phase settings to cover a specific angular sector.
	
	\begin{itemize}
		\item \textbf{Dynamically Configured RIS:} In this approach, a single $83 \times 83$ Connected-RIS is dynamically reconfigured using a register that stores all 52 codewords, each corresponding to a distinct beam steering direction. The FPGA sequentially activates these codewords to steer the beam as required. This configuration offers greater adaptability and occupies a significantly less physical space than the multi-panel alternative, making it well-suited for dynamic or space-constrained environments. However, it requires fast-switching circuitry to minimize potential service interruptions during the reconfiguration, which may otherwise result in coverage gaps. For instance, storing all 52 codewords requires approximately 1,075,684 bits for the fully controlled RIS with gain margin (6,889 elements $\times 3$ bits per element $\times 52$ codewords), and only 12,948 bits for the Connected-RIS (83 control units $\times 3$ bits $\times 52$ codewords). The RIS provides coverage and high signal quality for multiple users moving through the streets and inside buildings within a $160^{\circ}$ sector. We assume that the beam selection algorithm updates every 100 ms.
		\item \textbf{Fixed Multi-RIS Configuration:} In this configuration, 52 individual $83 \times 83$ Connected-RIS panels are deployed, each pre-configured with a fixed phase setting tailored to a specific angular sector. This eliminates the need for dynamic reconfiguration and the associated switching delays, enabling simultaneous coverage of all target directions and significantly simplifying the control circuitry. This advantage comes at the cost of increased physical deployment space and reduced adaptability to dynamic environments. This approach is particularly well-suited for wide-area high-density deployments, such as in smart city infrastructures. Each RIS panel is assigned a static codeword optimized for its designated coverage zone. As users move in the city, their devices connect to the most appropriate BS, while the fixed RIS panels continuously enhance the signal within their respective sectors. During peak hours, such as rush hours in the city center, this configuration ensures consistent high-quality coverage and robust connectivity for a dense concentration of UEs. This is particularly beneficial for user-centric services as well as machine-type communications, such as IoT applications operating in smart city environments.
	\end{itemize}	
	
	\begin{figure}[h!]
		\centering
		\hspace*{-0.2in}
		\includegraphics[width=4in,height=3.2in]{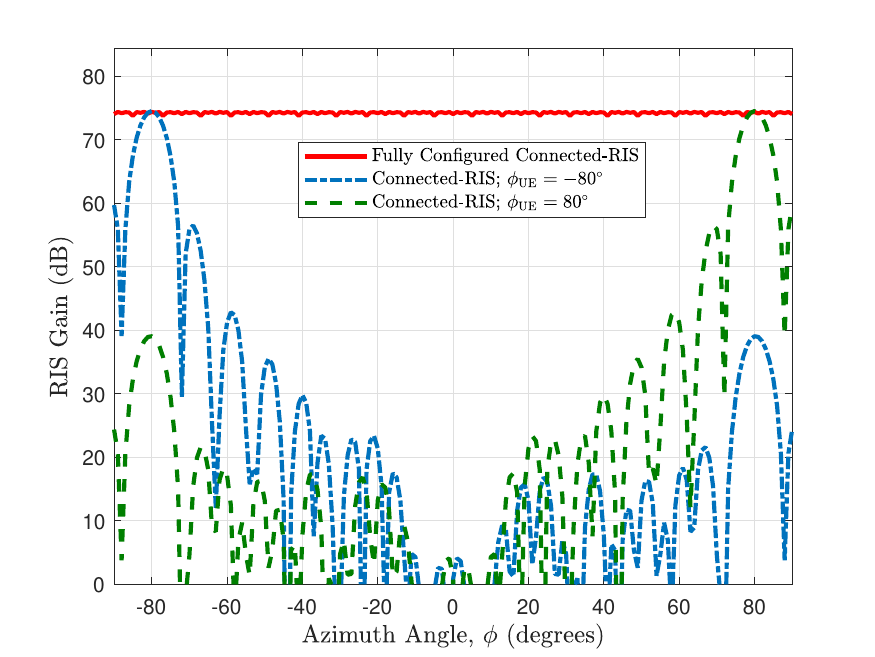}
		\caption{RIS gain versus the azimuth angle $\phi$ with multiple configurations to cover the region from $-80^{\circ}$ to $80^{\circ}$ in {\it Deployment Case 1}. Here, $\Delta G^{\rm dB}=6$ dB, $\psi_{\rm th}=31^{\circ}$.}
		\label{figconfig1}
	\end{figure}
	
	For both configuration approaches, the RIS gain performance is illustrated in Fig.~\ref{figconfig1} for representative steering angles ($\phi_{\rm UE}=-80^{\circ}$ and $\phi_{\rm UE}=80^{\circ}$).\footnote{The same behavior is observed when the gain margin is set to $\Delta G^{\rm dB}=3$ dB or when considering the other deployment cases ({\it Case 2} and {\it Case 3}). To reduce the number of figures, we present only the illustrative result shown in Fig.~\ref{figconfig1}.} The results indicate that the main lobes exhibit high gains, approaching 76 dB, while the side lobes remain significantly attenuated. By continuously configuring the RIS, full coverage of the said region can be obtained. This multi-beam strategy underscores the flexibility and efficiency of the Connected-RIS design in achieving wide-area coverage with reduced hardware complexity.
	\subsubsection{Power Consumption Evaluation}
	In this subsection, the power consumption of the dynamically configured RIS and the fixed multi-RIS configuration is discussed.
	\begin{itemize}
		\item \textbf{Dynamically Configured RIS:} For this setup, the total power consumption is influenced by multiple factors, including the RIS size, the configuration update rate, the type of switching elements, and the complexity of the control circuitry. Following the model in \cite{li2023enhan}, we assume a power requirement of 300 mW per configuration update for the update circuitry. Accordingly, the total power consumption of the Connected-RIS ($83\times 83$) is estimated at approximately 314.64 W, significantly lower than the 5628.78 W required by the fully controlled RIS with gain margin (6dB-RIS). For further comparison, the fully controlled RIS with minimum gain (min-gain RIS) under the same deployment conditions consumes about 2278.68 W. These findings underscore the energy efficiency of the Connected-RIS architecture, which achieves a substantial power reduction of 94.4\% and 86.20\% relative to the 6dB-RIS and the min-gain RIS designs, respectively. This highlights the practical viability of the Connected-RIS in energy-constrained environments, particularly for large-scale or real-time applications.
		
		\item \textbf{Fixed Multi-RIS Configuration:} Here, the power consumption remains relatively stable over time but is distributed across multiple RIS panels, with the total power determined by the cumulative consumption of each individual panel. For this setup, the total power consumption is estimated to be approximately 314.34 W for the Connected-RIS ($83\times 83$), compared to 5628.48 W for the 6dB-RIS and 2278.38 W for the min-gain RIS. Consistent with the dynamic configuration scenario, the Connected-RIS exhibits lower power consumption compared to the benchmark designs, reaffirming its suitability for scalable and power-sensitive deployments.
	\end{itemize}
	\section{Conclusion}
	This work introduced and validated a correlation-based RIS architecture, termed Connected-RIS, which effectively addresses the dual challenges of hardware simplification and energy consumption reduction in emerging RIS-aided wireless communication systems. By exploiting the correlation among the phase shifts values of the RIS, which is analyzed and confirmed for different deployments, the proposed design substantially reduces the number of independently controlled components, thereby minimizing the control signal requirements and simplifying the associated circuitry. Theoretical evaluations and extensive simulations across diverse deployment scenarios demonstrated that the Connected-RIS achieves significant reductions in both, the power consumption and the hardware complexity, while maintaining competitive beamforming gains and data rates. In comparison to the fully controlled RIS designs, the Connected-RIS offers a favorable balance between system performance and implementation cost, representing a critical advancement towards scalable and energy-aware RIS-enabled network architectures.		
	\bibliographystyle{IEEEtran}
	\bibliography{reference}
	
\end{document}